\documentclass[final,3p,nopreprintline]{elsarticle}

\usepackage[utf8]{inputenc}
\usepackage[T2A]{fontenc}
\usepackage[english]{babel}
\usepackage{amssymb}
\usepackage{amsmath}
\usepackage{hyperref}

\usepackage{algorithm}
\usepackage{algpseudocode}
\floatname{algorithm}{Algorithm}

\newcommand{\rot}{\mbox{\rm{rot\,}}}
 \newcommand{\diverg}{\mbox{\rm{div\,}}}

\journal{Journal of Computational Physics}

\begin{document}

\begin{frontmatter}

\title{Electromagnetic drift-kinetic particle-in-cell model with energy and charge conservation for studying finite-$\beta$ plasmas
 \tnoteref{t1}}
\tnotetext[t1]{The work is supported by Russian Science Foundation (grant № 25-11-00096) }

\author[nsu,binp]{O.P. Morozov}
\author[nsu,binp]{V.A. Kurshakov}
\author[nsu,binp]{I.V. Timofeev\corref{cor1}}
\ead{I.V.Timofeev@inp.nsk.su}

\affiliation[nsu]{organization={Novosibirsk State University},%
            addressline={Pirogov st., 1},
            city={Novosibirsk},
            postcode={630090},
            country={Russian Federation}}
\affiliation[binp]{organization={Budker Institute of Nuclear Physics SB RAS},%
            addressline={Lavrent'ev av., 11},
            city={Novosibirsk},
            postcode={630090},
            country={Russian Federation}}

\cortext[cor1]{Corresponding author}

\begin{abstract}

The paper proposes a generalization of the fully implicit energy- and charge-conserving electromagnetic particle-in-cell method to the case where the lightest type of plasma particle (electrons) is described in the drift-kinetic approximation. This allows us to remove the very strict time step limitation of this method requiring to resolve the gyrorotation of electrons. Since the drift-kinetic model is only applicable to light particles whose contribution to plasma polarization is small, this model can be further simplified by neglecting the electron polarization drift and including in Maxwell's equations, in addition to the current of gyrocenters, only the magnetization current. In order for such a hybrid model to retain conservative properties in finite-difference form, a method of self-consistent interpolation of $\nabla B$ in the mirror force from grid to particle and the magnetization vector from particle to grid is proposed. Unlike existing drift-kinetic models, this model is not limited to considering small perturbations near a given equilibrium, and therefore allows one to study the formation of plasma equilibria in regimes with a finite ratio of plasma to magnetic field pressure. Testing of the parallel code implemented in C++ using the PETSc library confirmed the fulfillment of the finite-difference laws of energy and charge conservation, as well as the ability of the model (in the drift-kinetic version for all types of particles) to correctly reproduce the diamagnetic effect and longitudinal ion-acoustic wave.
\end{abstract}


\end{frontmatter}

\section{Introduction}

The most complete description of fusion plasma in magnetic traps requires a joint solution of the Boltzmann kinetic equation and Maxwell's equations. The most efficient method for solving this system of equations in real 3D geometry is the particle-in-cell method. In this method, the plasma is represented as a set of finite-sized macroparticles, each of which moves during a time step under the influence of electromagnetic fields along the characteristics of the collisionless Vlasov equation, after which Monte Carlo algorithms determine the probability of additional changes in particle velocities due to their Coulomb collisions. The currents created by these particles are sources in Maxwell's equations and cause changes in the electric and magnetic fields specified on the spatial grid. Such a detailed kinetic description, however, is almost never used to model real fusion facilities. The point is that the smallest spatial and temporal scales (the Debye radius $r_D$ and the inverse plasma frequency $\omega_{pe}^{-1}$) that must be resolved in such a numerical model are many orders of magnitude smaller than the scales of plasma experiments. Moreover, the most easily implemented explicit PIC models typically do not conserve energy and therefore lack sufficient accuracy to describe such long-term processes. To solve this problem, a geometric PIC method \cite{Squire2012,Xiao2015} was proposed, which preserves the symplectic structure of the governing differential equations. In this case, finite-difference dynamic equations are derived from the variational principle for an already discretized action, which prevents accumulation of errors in the energy and other integrals of motion. Despite the fundamental solvability of the accuracy problem in the PIC method, such detailed PIC modeling of all particle types proves extremely resource-intensive (single calculations of a full-scale tokamak \cite{Xiao2021,XIAO_2021} and an open trap \cite{Zheng2021}, described in the literature, required a million computing cores and were limited to studying fairly rapid processes of microinstability development).

Thus, simplifying the physics of magnetic plasma confinement seems inevitable. The most obvious step in this direction is to neglect the rapid oscillations of electrons at the plasma frequency. However, in explicit PIC schemes, simply increasing the spatial and temporal steps ($\Delta x>r_D$ and $\Delta t>\omega_{pe}^{-1}$) leads to numerical instabilities. The possibility of not resolving such small scales appears only with implicit discretization of the exact equations of motion. Recently, several modifications of both the semi-implicit \cite{Lapenta2017,Lapenta2023} and fully implicit \cite{Markidis2011,Chen2011,Chen2014,Chen2023,Angus2023} PIC methods have been proposed, in which the energy conservation law is satisfied exactly even in finite-difference form. If, in addition, we use a special type of particle form-factor \cite{Chen2011,Chen2023} in the fully implicit approach and add a current correction stage \cite{Berendeev2024,Berendeev2024a} in the semi-implicit scheme, then we can also fulfill exactly the continuity equation, and hence Gauss's law. Examples of the application of this type of implicit PIC models to 3D modeling of magnetic traps are also still very few \cite{Ren2024,Park2026,Berendeev2026} and, as a rule, are limited to model problems with reduced plasma sizes and an underestimated ion-to-electron mass ratio. The modest gain that implicit models provide over explicit ones in modeling magnetic plasma confinement is explained by the need to resolve the electron cyclotron frequency $\Omega_e$ while maintaining a moderate excess of $\omega_{pe}$ over $\Omega_e$ in plasma experiments. Moreover, these models completely lose their advantage in open traps with strong magnetic mirrors, where the local electron cyclotron frequency can significantly exceed the plasma frequency ($\Omega_e\gg\omega_{pe}$).

The next step toward coarsening the kinetic description of plasma involves abandoning the resolution of the Larmor rotation of electrons. Indeed, if we are not interested in the problem of electron-cyclotron-resonance heating of a confined plasma, then a gyrokinetic description can be used for the electrons, in which the smooth motion of its guiding center is tracked instead of the rapidly oscillating motion of the particle itself. In this case, the time step can be significantly increased ($\Delta t\gg \Omega_e^{-1}$). Previously proposed hybrid PIC models \cite{Lee1978, Chen2009, Chen2019, Meng2025, Kurshakov2026}, in which electrons were described using the gyrokinetic or drift-kinetic approximation, were oriented towards modeling tokamaks with low relative plasma pressure ($\beta\ll 1$), and therefore either used expansions of perturbations of the particle distribution function and magnetic field over a small deviation from their equilibrium values, or did not contain electromagnetic effects at all. For modeling experiments in open traps, where the relative plasma pressure can no longer be considered small ($\beta< 0.6$), such perturbation theory is inapplicable. An alternative approach is to use the asymptotically preserving PIC method \cite{Ricketson2020,Ricketson2024}, which uses the same fully implicit discretization of the exact equations in both weak and strong magnetic fields. Although in a strong magnetic field (where $\Delta t\gg \Omega_e^{-1}$) the particle no longer follows an exact spiral trajectory, on average (asymptotically) it still correctly reproduces gyrokinetic drifts, with the exception of the magnetic gradient drift. A recipe for correcting gradient drift in a strong magnetic field without violating the law of energy conservation was given in \cite{Ricketson2020}, where an additional force is proposed to be introduced into the particle's equation of motion. An obstacle to the implementation of such an approach in the electromagnetic PIC model is the problem of calculating the magnetization current, which arises from the gyrorotation of particles and is incorrectly taken into account at large time steps.

Even simpler models suggest a transition to a hydrodynamic description of the electron fluid. The applicability of such a description to processes occurring in the tokamak plasma core appears entirely justified, but in SOL regions and open traps, where the plasma contacts a material wall, the validity of this approach is questionable. Recently, models assuming a Boltzmann distribution with a uniform and isotropic temperature for electrons in the ambipolar electric field \cite{Groenewald2025,Tran2025} have become popular for the numerical simulation of open systems. However, recent studies of electron transport in an open trap \cite{Glinskiy2024,Glinskiy2026} or a magnetic nozzle \cite{Tyushev2025} in the drift-kinetic approximation have shown that the weakly collisional nature of the plasma flow in expanders leads to a strong anisotropy of the electron distribution and a significant drop in their thermal energy along the magnetic field lines, which affects the longitudinal profiles of the electric potential and, ultimately, the stationary parameters of the confined plasma.

In this paper, we propose a new electromagnetic hybrid PIC model in which ions are described by the full Vlasov equation and the electron guiding centers move according to nonlinear drift kinetics, suitable for large $\beta$. Focusing primarily on modeling open traps, we will assume that the ${\bf E}\times{\bf B}$-drift velocity  is small compared to the thermal velocity of the electrons, and their polarization drift and associated current can be neglected. In this formulation of the problem, the law of energy conservation in differential form can be satisfied if the electron current in Maxwell's equations contains not only the current from the motion of the gyrocenters but also the magnetization current (diamagnetic current). To achieve the conservation of energy and charge in finite-difference form, we propose a fully implicit discretization of the drift equations, in which a method similar to Esirkepov's decomposition method \cite{Esirkepov2001} is used for self-consistent interpolation of the magnetic field gradient $\nabla B$ onto a particle in the mirror force and the magnetization vector ${\bf M}$ onto the Yee spatial grid. By analogy with the work of \cite{Chen2023}, this method can easily be generalized to the case of a particle flying through several grid cells.
The proposed drift-kinetic model is implemented as a module of the implicit \texttt{xpic} code solving the full Vlasov-Maxwell system \cite{Kurshakov2026}, but, at the moment, it is tested only in a drift-kinetic variant for all particle types.

The rest of the paper is structured as follows. In section \ref{s2}, the physical model is formulated: first, the equations of motion of the guiding centers, the diamagnetic current $\rot{\bf M}$ in Maxwell's equations, and the law of energy conservation following from them are presented in continuous (differential) form, and then a fully implicit discretization is proposed, in which the interpolation rules for $\nabla B$, $\rot{\bf B}$, and the magnetization ${\bf M}$ are consistent so as to conserve the energy of the system. Section \ref{s3} describes the software implementation: the time-stepping structure, two nested solvers --- an external solver for Maxwell's equations with Anderson acceleration and an internal Picard particle pusher, trajectory segmentation during multi-cell flybys, and parallelization using MPI and OpenMP. Section \ref{s4} tests the model with a series of numerical experiments, ranging from reproducing single-particle drifts and finite-difference conservation of energy and charge to diamagnetic equilibrium of a plasma sheet with finite $\beta$ and a traveling ion-acoustic wave. Section \ref{s5} summarizes the results and identifies future directions in developing this work.

\section{Description of the model}\label{s2}
\subsection{Physical model in continuous space and time}

To study the physics of plasma confinement in magnetic traps with a slowly varying non-uniform magnetic field, we propose to use a fully kinetic description for ions, which is provided by the Vlasov equation, and a drift-kinetic description for electrons, in which the real helical trajectories of particles are replaced by trajectories of their guiding centers.

With a fully kinetic description, the particle velocity distribution function
\begin{equation}
	f\left(t,{\bf r},{\bf v}\right)\propto \sum\limits_p \delta\left({\bf r}-{\bf R}_p(t)\right) \delta\left({\bf v}-{\bf v}_p(t)\right)
\end{equation}
obeys the Vlasov equation,
\begin{equation}
	\frac{\partial f}{\partial t}+{\bf v}\frac{\partial f}{{\bf r}}+\frac{q}{m}\left({\bf E}+\left[{\bf v}\times {\bf B}\right]\right)\frac{\partial f}{\partial {\bf v}}=0,
\end{equation}
 the characteristics of which are the equations of motion of individual particles:
\begin{align}
& \frac{d {\bf R}_p}{dt}={\bf v}_p, \\
	&m_p \frac{d{\bf v}_p}{dt}=q_p \left({\bf E}_p+\left[{\bf v}_p\times {\bf B}_p\right]\right),
\end{align}
where ${\bf R}_p(t)$ is the radius vector of the particle, ${\bf v}_p(t)$ is its velocity vector, ${\bf E}_p={\bf E}(t, {\bf R}_p(t))$ and ${\bf B}_p={\bf B}(t, {\bf R}_p(t))$ are the electric and magnetic fields acting on the particle, $m_p$ and $q_p$ are the mass and charge of the particle. The following system of units is used here: charge is measured in units of the elementary charge of an electron $e$, mass --- in electron masses $m_e$, time --- in $\omega_{pe}^{-1}$, where $\omega_{pe}=\sqrt{4\pi e^2 n_0/m_e}$ is the plasma frequency calculated for the characteristic density $n_0$, velocity --- in units of the speed of light $c$, distance --- in $c/\omega_{pe}$, electromagnetic fields --- in $m_e c \omega_{pe}/e$. The self-consistent evolution of the electromagnetic field is determined by Maxwell's equations, in which the role of the source is played by the current
\begin{equation}
	{\bf J}(t,{\bf r})=\sum\limits_p q_p {\bf v}_p \delta\left({\bf r}-{\bf R}_p\right),
\end{equation}
which is measured in units of $en_0c$.

In the problem of magnetic plasma confinement, such a detailed description seems redundant for electrons. Indeed, given that the electric and magnetic fields  in magnetic traps vary slowly compared to the electron's cyclotron rotation period, and the characteristic scales of their spatial inhomogeneity are significantly larger than the electron gyroradius, instead of the exact equations of motion of real particles, we can use simplified equations of motion for electrons' guiding centers. Neglecting the polarization drift of electrons and assuming that the electric drift velocity is small compared to their thermal velocity, the drift motion can be described by the following equations:
\begin{align}
	&\frac{d{\bf R}_p}{dt}=v_{\|p}\frac{{\bf B}_p^{\ast}}{B_p^{\ast}}+\left[\left({\bf E}_p-\frac{\mu_p}{q_p}\nabla B_p\right)\times \frac{{\bf b}_p}{B_p}\right], \\
	&m_p \frac{d v_{\|p}}{dt}=\left(q_p {\bf E}_p- \mu_p \nabla B_p\right)\frac{{\bf \dot{R}}_p}{v_{\|p}}, \\
	&{\bf B}_p^{\ast}={\bf B}_p +\frac{m_p v_{\|p}}{q_p} \left(\rot {\bf b}\right)_p, \quad B_p^{\ast}={\bf b}_p{\bf B}_p^{\ast},
\end{align}
where ${\bf R}_p$ and $v_{\|p}$ are now the radius vector and longitudinal velocity of the particle's guiding center, $B_p=|{\bf B}_p|$ and ${\bf b}_p={\bf B}_p/B_p$ are the modulus and unit vector along the magnetic field direction, $(\nabla B)_p$ is the gradient of the magnetic field modulus at the particle's location, and $\mu_p=m_p v_{\bot p}^2/(2B_p)$ is the magnetic moment of the Larmor circle, which is independent of time and the particle's position, since it is an adiabatic invariant in the approximation used. From the given equations it follows that the kinetic energy of plasma particles in arbitrary variable fields changes according to the following law:
\begin{equation}\label{kin}
	\frac{d}{dt}\sum\limits_p\left(m_p \frac{v_{\|p}^2}{2}+\mu_p B_p\right)=\sum\limits_p q_p {\bf E}_p {\bf \dot{R}}_p +\sum\limits_p\mu_p \left.\frac{\partial B(t,{\bf r})}{\partial t}\right|_{{\bf r}={\bf R}_p}.
\end{equation}
Here we take into account that the change in the modulus of the magnetic field occurs not only due to its local change, but also due to the displacement of the particle to another spatial point:
\begin{equation}
	\frac{dB(t,{\bf R}_p(t))}{dt}=\frac{\partial B}{\partial t}+\left({\bf \dot{R}}_p\nabla\right)B.
\end{equation}

The evolution of electromagnetic fields is described by Maxwell's equations
\begin{align}
	&\rot {\bf B}=\frac{\partial {\bf E}}{\partial t}+{\bf J} +\rot {\bf M}, \\
	&\rot {\bf E}=-\frac{\partial {\bf B}}{\partial t},
\end{align}
where, along with the current from the gyrocenters of electrons
\begin{equation}
	{\bf J}=\sum\limits_{p} q_p {\bf \dot{R}}_p \delta\left({\bf r}-{\bf R}_p\right),
	\end{equation}
we also include the diamagnetic current $\rot {\bf M}$ created by their cyclotron rotation, where the magnetization vector is determined by the expression
\begin{equation}
	{\bf M}=-\sum\limits_p \mu_p {\bf b}_p \delta\left({\bf r}-{\bf R}_p\right).
\end{equation}
From the presented Maxwell equations we conclude that the energy of the electromagnetic field changes over time as follows
\begin{equation}
	\frac{\partial}{\partial t}\int dV\left(\frac{|{\bf E}|^2}{2}+\frac{|{\bf B}|^2}{2}\right)=-\int {\bf E}{\bf J} dV + \int {\bf M} \frac{\partial {\bf B}}{\partial t} dV+\oint \left[{\bf E}\times {\bf H}\right] d{\bf S},
\end{equation}
where ${\bf H}={\bf B}-{\bf M}$. Substituting here the definitions of ${\bf J}$ and ${\bf M}$, we obtain
\begin{equation}\label{en}
	\frac{\partial}{\partial t}\int dV\left(\frac{|{\bf E}|^2}{2}+\frac{|{\bf B}|^2}{2}\right)=-\sum\limits_p q_p {\bf \dot{R}}_p {\bf E}_p - \sum\limits_p \mu_p\int {\bf b} \frac{\partial {\bf B}}{\partial t} \delta\left({\bf r}-{\bf R}_p\right) dV+\oint \left[{\bf E}\times {\bf H}\right] d{\bf S}.
\end{equation}
Since
\begin{equation}
	{\bf b}\frac{\partial {\bf B}}{\partial t}=\frac{\partial \left({\bf B}\cdot{\bf b}\right)}{\partial t}-{\bf B}\frac{\partial {\bf b}}{\partial t}=\frac{\partial B}{\partial t},
\end{equation}
the first two terms on the right-hand side of (\ref{en}) coincide with the right-hand side of the equation (\ref{kin}) and describe (with the opposite sign) the change in the kinetic energy of the particles.  Thus, in a system where the motion of particles is reduced to the drift motion of point gyrocenters, the law of energy conservation takes the form:
\begin{equation}
	\frac{\partial}{\partial t}\int dV\left(\frac{|{\bf E}|^2}{2}+\frac{|{\bf B}|^2}{2}\right)+\frac{d}{dt}\sum\limits_p \left(m_p \frac{v_{\|p}^2}{2}+\mu_p B_p\right) =\oint \left[{\bf E}\times {\bf H}\right] d{\bf S}.
\end{equation}

\subsection{Fully implicit discretization}\label{s22}

To discretize the exact equations of ion motion, we will use the fully implicit Crank-Nicolson scheme:
\begin{align}
  &\frac{{\bf R}_p^{n+1}-{\bf R}_p^{n}}{\tau}={\bf v}_p^{n+1/2}, \\
	&m_p\frac{{\bf v}_p^{n+1}-{\bf v}_p^{n}}{\tau}=q_p \left({\bf E}_p^{n+1/2}+\left[{\bf v}_p^{n+1/2}\times{{\bf B}_p^{n+1/2}}\right]\right),
\end{align}
where $\tau$ is the time step, and all quantities in the middle of the time step at $t=(n+1/2)\tau$ are calculated using the rule $f^{n+1/2}=(f^{n+1}+f^n)/2$. For space discretization, we use the standard Yee grid in Cartesian coordinates. As in the paper \cite{Chen2023}, the interpolation of the grid  electric field ${\bf E}_g$ onto the particle is performed using the same asymmetric kernel ${\bf \mathcal{S}}_{gp}$ that is used to interpolate the current from the particle to the grid:
\begin{align}
	{\bf E}_p^{n+1/2}=&\sum\limits_{g} {\bf E}_g^{n+1/2} {\bf \mathcal{S}}_{gp}\left({\bf r}_g,{\bf R}_p^{n+1},{\bf R}_p^{n}\right)=\nonumber \\
	&={\bf i}\sum\limits_{i,j,k} E_{xi+1/2,j,k}^{n+1/2} \mathcal{S}_{xx}+{\bf j}\sum\limits_{i,j,k} E_{yi,j+1/2,k}^{n+1/2} \mathcal{S}_{yy}+{\bf k}\sum\limits_{i,j,k} E_{zi,j,k+1/2}^{n+1/2} \mathcal{S}_{zz}. \label{ef}
\end{align}
It is evident that ${\bf \mathcal{S}}_{gp}$ is a tensor in which only the diagonal components $\mathcal{S}_{ii}$ are nonzero. The form of this function is chosen such that the interpolations of the particle density and its current density exactly satisfy the continuity equation. Applying Esirkepov's density decomposition method to each segment $s$ of the straight trajectory between ${\bf R}_p^n$ and ${\bf R}_p^{n+1}$, lying in a separate cell, and choosing the parabolic kernel $S_2$ for the density, we find that in the direction of the shift along each coordinate, the order of smoothness of the current interpolation decreases by one and is described by the linear kernel $S_1$:
\begin{align}
	\mathcal{S}_{xx}=&\sum\limits_{s\in n}\frac{X_p^{s+1}-X_p^s}{X_p^{n+1}-X_p^n} S_1(x_{i+1/2}-X_p^{s+1/2}) S_{22}^{yz}[j,k], \\
	\mathcal{S}_{yy}=&\sum\limits_{s\in n}\frac{Y_p^{s+1}-Y_p^s}{Y_p^{n+1}-Y_p^n} S_1(y_{j+1/2}-Y_p^{s+1/2}) S_{22}^{xz}[i,k], \\
	\mathcal{S}_{zz}=&\sum\limits_{s\in n}\frac{Z_p^{s+1}-Z_p^s}{Z_p^{n+1}-Z_p^n} S_1(z_{k+1/2}-Z_p^{s+1/2}) S_{22}^{xy}[i,j],
\end{align}
where
\begin{align}
	S_{22}^{yz}[j,k]=\frac{1}{3}&\left[S_2\left(y_j-Y_p^s\right)S_2\left(z_k-Z_p^s\right)+\frac{S_2\left(y_j-Y_p^s\right)S_2\left(z_k-Z_p^{s+1}\right)}{2} \right.  \nonumber \\
	&\left. +\frac{S_2\left(y_j-Y_p^{s+1}\right)S_2\left(z_k-Z_p^{s}\right)}{2}+S_2\left(y_j-Y_p^{s+1}\right)S_2\left(z_k-Z_p^{s+1}\right)\right] \\
	S_{22}^{xz}[i,k]=\frac{1}{3}&\left[S_2\left(x_i-X_p^s\right)S_2\left(z_k-Z_p^s\right)+\frac{S_2\left(x_i-X_p^s\right)S_2\left(z_k-Z_p^{s+1}\right)}{2} \right.  \nonumber \\
	&\left. +\frac{S_2\left(x_i-X_p^{s+1}\right)S_2\left(z_k-Z_p^{s}\right)}{2}+S_2\left(x_i-X_p^{s+1}\right)S_2\left(z_k-Z_p^{s+1}\right)\right]  \\
	S_{22}^{xy}[i,j]=\frac{1}{3}&\left[S_2\left(y_j-Y_p^s\right)S_2\left(x_i-X_p^s\right)+\frac{S_2\left(y_j-Y_p^s\right)S_2\left(x_i-X_p^{s+1}\right)}{2} \right.  \nonumber \\
	&\left. +\frac{S_2\left(y_j-Y_p^{s+1}\right)S_2\left(x_i-X_p^{s}\right)}{2}+S_2\left(y_j-Y_p^{s+1}\right)S_2\left(x_i-X_p^{s+1}\right)\right].
\end{align}
Summation over $s$ in this formula actually means averaging the electric field on a particle along its trajectory over one time step. Using the same interpolation function for both the electric field and current density generated by the particles
\begin{equation}\label{jf}
	{\bf J}_g^{n+1/2}=\sum\limits_p q_p \frac{{\bf R}_p^{n+1}-{\bf R}_p^{n}}{\tau} {\bf \mathcal{S}}_{gp}\left({\bf r}_g,{\bf R}_p^{n+1},{\bf R}_p^{n}\right)
\end{equation}
allows for the exact energy conservation even in finite-difference form. Since the continuity equation is satisfied exactly for each particle,
\begin{equation}\label{cont}
	\frac{\rho_g^{n+1}-\rho_g^{n}}{\tau}+\left(\diverg {\bf J}\right)_g^{n+1/2}=0, \quad \rho_g^{n}=\sum\limits_p q_p S_2\left({\bf r}_g-{\bf R}_p^{n}\right),
\end{equation}
the Gauss's law $\left(\diverg {\bf E}^n\right)_g=\rho_g^n$, satisfied at the initial moment of time, remains valid at all subsequent steps. The magnetization current does not violate this property, since the finite-difference identity $\diverg\rot\equiv 0$ is fulfilled exactly on the Yee grid.

Since the magnetic field does no work, any function can be used to interpolate its grid values onto a particle. In particular, the parabolic kernel $S_2$ can be chosen:
\begin{equation}\label{bf}
	{\bf B}_p^n=\sum\limits_g {\bf B}_g^n S_2({\bf r}_g-{\bf R}_p^n).
\end{equation}

A finite-difference approximation of guiding center equations for electrons will also be carried out using a fully implicit approach:
\begin{align}
	&\frac{{\bf R}_p^{n+1}-{\bf R}_p^{n}}{\tau}=v_{\|p}^{n+1/2}\frac{{\bf B}_p^{\ast n+1/2}}{B_p^{\ast n+1/2}}+\left[\left(q_p{\bf E}_p^{n+1/2}-\mu_p\left(\nabla B\right)_p^{n+1/2}\right)\times \frac{{\bf b}_p^{n+1/2}}{q_p B_p^{n+1/2}}\right]\equiv {\bf V}_p^{n+1/2}, \label{drift_r}\\
	&m_p \frac{v_{\|p}^{n+1}-v_{\|p}^{n}}{\tau}= \left(q_p{\bf E}_p^{n+1/2}-\mu_p\left(\nabla B\right)_p^{n+1/2}\right) \frac{{\bf B}_p^{\ast n+1/2}}{B_p^{\ast n+1/2}}\equiv F_p^{n+1/2},        \label{vpar}
\end{align}
where
\begin{align}
	&{\bf B}_p^{\ast n+1/2}=B_p^{n+1/2}{\bf \hat{b}}_p^{n+1/2}+\frac{m_p v_{\|p}^{n+1/2}}{q_p} \left(\rot{\bf b}\right)_p^{n+1/2}, \quad B_p^{\ast n+1/2}={\bf \hat{b}}_p^{n+1/2}{\bf B}_p^{\ast n+1/2}, \label{bstar}\\
	&\left(\rot{\bf b}\right)_p^{n+1/2}=\left[{\bf b}_p^{n+1/2}\times \frac{\left(\nabla B\right)_p^{n+1/2}}{B_p^{n+1/2}}\right]+ \frac{\left(\rot {\bf B}\right)_p^{n+1/2}}{B_p^{n+1/2}}.
\end{align}
Here the quantities $B_p^{n+1/2}$, ${\bf b}_p^{n+1/2}$ and ${\bf \hat{b}}_p^{n+1/2}$ at the half-integer step are constructed according to the rules:
\begin{equation}\label{bhalf}
	B_p^{n+1/2}=\frac{\left|{\bf B}_p^{n+1}\right|+\left|{\bf B}_p^{n}\right|}{2}, \quad {\bf b}_p^{n+1/2}=\frac{{\bf b}_p^{n+1}+{\bf b}_p^{n}}{2}, \quad {\bf \hat{b}}_p^{n+1/2}=\frac{{\bf b}_p^{n+1/2}}{\left|{\bf b}_p^{n+1/2}\right|},
\end{equation}
where ${\bf b}_p^{n}={\bf B}_p^{n}/\left|{\bf B}_p^{n}\right|$. The half-sum of two unit vectors is not itself a unit vector, so below we must carefully distinguish between ${\bf b}_p^{n+1/2}$ and the normalized vector ${\bf \hat{b}}_p^{n+1/2}$. The method for calculating the fields ${\bf E}_p^{n+1/2}$ and ${\bf B}_p^{n}$, ${\bf B}_p^{n+1}$ on a particle remains the same (\ref{ef}) and (\ref{bf}), and the interpolation rule for $\left(\nabla B\right)_p^{n+1/2}$ and $\left(\rot {\bf B}\right)_p^{n+1/2}$ will be established later from the requirement of conservation of the system's energy.

Maxwell's finite difference equations can be written in the form:
\begin{align}
	&\left(\rot {\bf H}\right)_g^{n+1/2}=\frac{{\bf E}_g^{n+1}-{\bf E}_g^{n}}{\tau}+{\bf J}_g^{n+1/2}, \\
	&\left(\rot {\bf E}\right)_g^{n+1/2}=-\frac{{\bf B}_g^{n+1}-{\bf B}_g^{n}}{\tau},
\end{align}
where the grid current is calculated using the rule (\ref{jf}) not only for Vlasov ions, but also for electron gyrocenters. Multiplying the first equation by ${\bf E}_g^{n+1/2}$, the second by ${\bf B}_g^{n+1/2}$, adding them up, and summing over all grid nodes, we obtain the growth rate of the electromagnetic field energy
\begin{align}
	\frac{1}{\tau}\sum\limits_g &\left(\frac{\left|{\bf E}_g^{n+1}\right|^2-\left|{\bf E}_g^{n}\right|^2}{2}+\frac{\left|{\bf B}_g^{n+1}\right|^2-\left|{\bf B}_g^{n}\right|^2}{2}\right)= \nonumber \\
	&-\sum\limits_g {\bf E}_g^{n+1/2}{\bf J}_g^{n+1/2}+\sum\limits_g{\bf M}_g^{n+1/2}\frac{{\bf B}_g^{n+1}-{\bf B}_g^{n}}{\tau}+\sum\limits_g\left(\diverg \left[{\bf E}\times{\bf H}\right]\right)_g^{n+1/2}.
\end{align}
This equation will acquire the meaning of the law of total energy conservation only if the first two terms on its right-hand side are equal to the change in the kinetic energy of all plasma particles with the opposite sign, that is, our finite-difference scheme must ensure equality
\begin{equation}\label{kin_max}
	\frac{\Delta W_K}{\tau}=\sum\limits_g {\bf E}_g^{n+1/2}{\bf J}_g^{n+1/2}-\sum\limits_g{\bf M}_g^{n+1/2}\frac{{\bf B}_g^{n+1}-{\bf B}_g^{n}}{\tau}.
\end{equation}
Thus, the problem is reduced to finding a method for interpolating the magnetization vector and the magnetic field gradient, in which the above expression can be transformed to the form
\begin{equation}\label{kin2}
	\frac{\Delta W_K}{\tau}=\frac{1}{\tau}\sum\limits_p\left(m_p \frac{(v_{\|p}^{n+1})^2-(v_{\|p}^{n})^2}{2}+\mu_p \left(B_p^{n+1}-B_p^n\right)\right).
\end{equation}
Since the vector ${\bf M}$ enters Maxwell's equations only at the half-integer time step $n+1/2$, the logic of implicit discretization suggests the following interpolation method:
\begin{equation}
	{\bf M}_g^{n+1/2}=-\sum\limits_p \mu_p \frac{{\bf b}_p^{n+1/2}}{|{\bf b}_p^{n+1/2}|^2} \frac{S_2^{n+1}+S_2^n}{2},\label{magn}
\end{equation}
where $S_2^n$ denotes the same parabolic kernel $S_2({\bf r}_g-{\bf R}_p^n)$, which is used to interpolate the magnetic field vector from the grid to the particle, and the factor $\left|{\bf b}_p^{n+1/2}\right|^2$ takes into account the difference from unity of the length of the vector ${\bf b}_p^{n+1/2}$. Substituting the expressions for the current ${\bf J}_g^{n+1/2}$ and magnetization ${\bf M}_g^{n+1/2}$ into the formula (\ref{kin_max}), and also changing the order of summation over particles and over grid nodes, we obtain
\begin{align}\label{kin1}
	\frac{\Delta W_K}{\tau}=\sum\limits_p q_p {\bf V}_p^{n+1/2} {\bf E}_p^{n+1/2}+ \sum\limits_p \mu_p \frac{{\bf b}_p^{n+1/2}}{|{\bf b}_p^{n+1/2}|^2} \sum\limits_g \frac{({\bf B}_g^{n+1}-{\bf B}_g^{n})}{\tau} \frac{(S_2^{n+1}+S_2^n)}{2}.
\end{align}
Using equality
\begin{equation}
	\frac{({\bf B}_g^{n+1}-{\bf B}_g^{n})}{\tau} \frac{(S_2^{n+1}+S_2^n)}{2}=\frac{{\bf B}_g^{n+1} S_2^{n+1}-{\bf B}_g^{n} S_2^{n}}{\tau}-{\bf B}_g^{n+1/2} \frac{(S_2^{n+1}-S_2^n)}{\tau}
\end{equation}
the expression (\ref{kin1}) is reduced to the form
\begin{align}
	\frac{\Delta W_K}{\tau}=\sum\limits_p q_p {\bf V}_p^{n+1/2} {\bf E}_p^{n+1/2}+ \sum\limits_p \mu_p \frac{{\bf b}_p^{n+1/2}}{|{\bf b}_p^{n+1/2}|^2}\left(\frac{{\bf B}_p^{n+1}-{\bf B}_p^{n}}{\tau}-\sum\limits_g {\bf B}_g^{n+1/2} \frac{(S_2^{n+1}-S_2^n)}{\tau}\right).
\end{align}
Considering that
\begin{equation}
	\frac{{\bf B}_p^{n+1}-{\bf B}_p^{n}}{\tau}=B_p^{n+1/2}\frac{{\bf b}_p^{n+1}-{\bf b}_p^{n}}{\tau}+{\bf b}_p^{n+1/2}\frac{B_p^{n+1}-B_p^n}{\tau},
\end{equation}
and the vector ${\bf b}_p^{n+1/2}$ is orthogonal to the vector ${\bf b}_p^{n+1}-{\bf b}_p^{n}$, we obtain
\begin{align}
	\frac{\Delta W_K}{\tau}=\sum\limits_p\left[ q_p {\bf V}_p^{n+1/2} {\bf E}_p^{n+1/2}+ \mu_p \frac{B_p^{n+1}-B_p^n}{\tau} - \mu_p \frac{{\bf b}_p^{n+1/2}}{|{\bf b}_p^{n+1/2}|^2}\sum\limits_g {\bf B}_g^{n+1/2} \frac{(S_2^{n+1}-S_2^n)}{\tau} \right] .
\end{align}
The work of the electric field in moving the gyrocenter of a particle can be expressed through the change in its longitudinal energy. Indeed, multiplying equation (\ref{vpar}) by $v_{\|p}^{n+1/2}$ and using the fact that the transverse term in (\ref{drift_r}) is orthogonal to the vector $q_p{\bf E}_p^{n+1/2}-\mu_p\left(\nabla B\right)_p^{n+1/2}$, the first term in square brackets can be represented as
\begin{equation}
	q_p {\bf V}_p^{n+1/2} {\bf E}_p^{n+1/2}= \frac{m_p}{2\tau}\left( (v_{\|p}^{n+1})^2-(v_{\|p}^{n})^2\right) + \mu_p {\bf V}_p^{n+1/2}\cdot \left(\nabla B\right)_p^{n+1/2}.
\end{equation}
It can be seen that the final expression for the rate of change of kinetic energy
\begin{align}
	\frac{\Delta W_K}{\tau}=&\frac{1}{\tau}\sum\limits_p\left(m_p \frac{(v_{\|p}^{n+1})^2-(v_{\|p}^{n})^2}{2}+\mu_p \left(B_p^{n+1}-B_p^n\right)\right)+\nonumber \\
	&+ \sum\limits_p \mu_p \left[{\bf V}_p^{n+1/2}\cdot \left(\nabla B\right)_p^{n+1/2}-\frac{{\bf b}_p^{n+1/2}}{|{\bf b}_p^{n+1/2}|^2}\sum\limits_g {\bf B}_g^{n+1/2} \frac{(S_2^{n+1}-S_2^n)}{\tau}\right]
\end{align}
reduces to formula (\ref{kin2}) if the last two terms exactly compensate each other. This means that the law of global energy conservation in the chosen numerical scheme can be fulfilled exactly even in finite-difference form, if the equation
\begin{equation}\label{cond}
	{\bf V}_p^{n+1/2}\cdot \left(\nabla B\right)_p^{n+1/2}=\frac{{\bf b}_p^{n+1/2}}{|{\bf b}_p^{n+1/2}|^2}\sum\limits_g {\bf B}_g^{n+1/2} \frac{(S_2^{n+1}-S_2^n)}{\tau}
\end{equation}
is considered as a rule for interpolating the gradient of the magnetic field modulus onto a particle. At first glance, it seems impossible to find the three components of the vector $\nabla B$ using a single scalar equation, but the problem that arises here turns out to be similar to the problem of finding three current components from a single scalar continuity equation. Indeed, applying Esirkepov's decomposition method to the difference in form factors
\begin{align}
	&S_2\left({\bf r}_g-{\bf R}_p^{n+1}\right)-S_2\left({\bf r}_g-{\bf R}_p^{n}\right)=\left[S_2\left(x_g-X_p^{n+1}\right)-S_2\left(x_g-X_p^{n}\right)\right]S_{22}^{yz}+\nonumber \\
	&+\left[S_2\left(y_g-Y_p^{n+1}\right)-S_2\left(y_g-Y_p^{n}\right)\right]S_{22}^{xz}+\left[S_2\left(z_g-Z_p^{n+1}\right)-S_2\left(z_g-Z_p^{n}\right)\right]S_{22}^{xy}
\end{align}
and taking advantage of the fact that differences of the form
\begin{equation}
	S_2\left(x_g-X_p^{n+1}\right)-S_2\left(x_g-X_p^{n}\right)\propto X_p^{n+1}-X_p^n=V_{px}^{n+1/2}\tau
\end{equation}
are proportional to the corresponding components of the velocity vector ${\bf V}_p^{n+1/2}$ (if the trajectory $x\in\left[X_p^n;X_p^{n+1}\right]$ lies within a single cell), we represent the right-hand side of equation (\ref{cond}) as a sum of terms, which can be interpreted as a scalar product of the form $({\bf V}_p^{n+1/2}\cdot ...)$. Thus, to satisfy equation (\ref{cond}), the $x$-component of the gradient of the magnetic field modulus on the particle must be found from the equation
\begin{align}
	\left|{\bf b}_p^{n+1/2}\right|^2 V_{px}^{n+1/2}&\left(\nabla_x B\right)_p^{n+1/2}=\nonumber \\
	& b_{px}^{n+1/2}\sum\limits_{i,j,k} (B_x)_{i,j+1/2,k+1/2}^{n+1/2} \frac{S_2\left(x_i-X_p^{n+1}\right)-S_2\left(x_i-X_p^{n}\right)}{\tau} \left(S_{22}^{yz}\right)_{j+1/2,k+1/2}+\nonumber \\
	& b_{py}^{n+1/2}\sum\limits_{i,j,k} (B_y)_{i+1/2,j,k+1/2}^{n+1/2} \frac{S_2\left(x_{i+1/2}-X_p^{n+1}\right)-S_2\left(x_{i+1/2}-X_p^{n}\right)}{\tau} \left(S_{22}^{yz}\right)_{j,k+1/2}+\nonumber \\
	& b_{pz}^{n+1/2}\sum\limits_{i,j,k} (B_z)_{i+1/2,j+1/2,k}^{n+1/2} \frac{S_2\left(x_{i+1/2}-X_p^{n+1}\right)-S_2\left(x_{i+1/2}-X_p^{n}\right)}{\tau} \left(S_{22}^{yz}\right)_{j+1/2,k}. \label{gr}
\end{align}
To use the same segmentation of the trajectory  that is applied when we calculate the interpolation function ${\bf \mathcal{S}}_{gp}$ for the electric field and current, we must assume that $X_p^n$ and $X_p^{n+1}$ are located in the volume between the half-integer nodes of the spatial grid $i_p-1/2$ and $i_{p}+1/2$, where $x_p=i_p\Delta x$ is the coordinate of the integer node near which the particle's trajectory lies. In this case, the following equalities hold:
\begin{align}
	\sum\limits_{i} &(B_x)_i^{n+1/2} \frac{S_2\left(x_i-X_p^{n+1}\right)-S_2\left(x_i-X_p^{n}\right)}{\tau}=\nonumber \\
	&=V_{px}^{n+1/2} \sum\limits_{i} \frac{(B_{x})_{i+1}^{n+1/2}-(B_{x})_{i}^{n+1/2}}{\Delta x} S_1\left(x_{i+1/2}-X_p^{n+1/2}\right), \\
	\sum\limits_{i} &(B_{y,z})_{i+1/2}^{n+1/2} \frac{S_2\left(x_{i+1/2}-X_p^{n+1}\right)-S_2\left(x_{i+1/2}-X_p^{n}\right)}{\tau}=V_{px}^{n+1/2} \sum\limits_{i} \frac{(B_{y,z})_{i+1/2}^{n+1/2}-(B_{y,z})_{i-1/2}^{n+1/2}}{\Delta x}\times\nonumber \\
	&\times\left[\frac{X_p^{n+1}-x_p}{X_p^{n+1}-X_p^n} S_1\left(x_{i}-\frac{x_p+X_p^{n+1}}{2}\right)+\frac{x_p-X_p^{n}}{X_p^{n+1}-X_p^n} S_1\left(x_{i}-\frac{x_p+X_p^{n}}{2}\right)\right].
\end{align}
Due to the relative shift of various field components on the Yee grid, the particle's trajectory in some cases must be additionally subdivided at the position of the nearest integer node $x_p$, which results in the two terms in the second equality. Substituting these equalities into formula (\ref{gr}) and generalizing to the case of a particle passing through several cells, we obtain the following rule for interpolating the gradients of the grid magnetic field on a particle:
\begin{align}
	\left(\nabla_x B\right)_p^{n+1/2}=&\frac{b_{px}^{n+1/2}}{\left|{\bf b}_p^{n+1/2}\right|^2}\sum\limits_{i,j,k} \frac{(B_x)_{i+1,j+1/2,k+1/2}^{n+1/2}-(B_x)_{i,j+1/2,k+1/2}^{n+1/2}}{\Delta x} S_{xx}^{B}+\nonumber \\
	+&\frac{b_{py}^{n+1/2}}{\left|{\bf b}_p^{n+1/2}\right|^2}\sum\limits_{i,j,k} \frac{(B_y)_{i+1/2,j,k+1/2}^{n+1/2}-(B_y)_{i-1/2,j,k+1/2}^{n+1/2}}{\Delta x} S_{xy}^B+\nonumber \\
	+&\frac{b_{pz}^{n+1/2}}{\left|{\bf b}_p^{n+1/2}\right|^2}\sum\limits_{i,j,k} \frac{(B_z)_{i+1/2,j+1/2,k}^{n+1/2}-(B_z)_{i-1/2,j+1/2,k}^{n+1/2}}{\Delta x} S_{xz}^B,\label{gradB}
\end{align}
where
\begin{align}
	S_{xx}^{B}=\sum\limits_{s\in n} &\frac{X_p^{s+1}-X_p^s}{X_p^{n+1}-X_p^n} S_1\left(x_{i+1/2}-X_p^{s+1/2}\right) S_{22}^{yz}[j+1/2,k+1/2], \\
	S_{xy}^{B}=\sum\limits_{s\in n} &\left[\frac{X_p^{s+1}-x_s}{X_p^{n+1}-X_p^n} S_1\left(x_{i}-\frac{x_s+X_p^{s+1}}{2}\right)+\frac{x_s-X_p^{s}}{X_p^{n+1}-X_p^n} S_1\left(x_{i}-\frac{x_s+X_p^{s}}{2}\right)\right]S_{22}^{yz}[j,k+1/2], \\
	S_{xz}^{B}=\sum\limits_{s\in n} &\left[\frac{X_p^{s+1}-x_s}{X_p^{n+1}-X_p^n} S_1\left(x_{i}-\frac{x_s+X_p^{s+1}}{2}\right)+\frac{x_s-X_p^{s}}{X_p^{n+1}-X_p^n} S_1\left(x_{i}-\frac{x_s+X_p^{s}}{2}\right)\right]S_{22}^{yz}[j+1/2,k].
\end{align}
As before, the particle trajectory at a time step is divided here into segments, each of which lies in a volume constructed around the node $(i_s,j_s,k_s)$. $(X_p^{s},Y_p^s,Z_p^s)$ and $(X_p^{s+1},Y_p^{s+1},Z_p^{s+1})$ are the coordinates of the beginning and end of the segment with number $s$, $(x_s,y_s,z_s)=(i_s\Delta x, j_s \Delta y, k_s \Delta z)$ is the coordinate of the node of the cell that we consider. If a less smooth form factor $S_1$ is used for ${\bf B}_p$ and ${\bf M}_g$, then the functions $S_0$ (NGP kernel) and $S_1$ will participate in the interpolation of the magnetic field gradient.

Expressions for other components of the magnetic field gradient are obtained similarly:
\begin{align}
	\left(\nabla_y B\right)_p^{n+1/2}=&\frac{b_{px}^{n+1/2}}{\left|{\bf b}_p^{n+1/2}\right|^2}\sum\limits_{i,j,k} \frac{(B_x)_{i,j+1/2,k+1/2}^{n+1/2}-(B_x)_{i,j-1/2,k+1/2}^{n+1/2}}{\Delta y} S_{yx}^{B}+\nonumber \\
	+&\frac{b_{py}^{n+1/2}}{\left|{\bf b}_p^{n+1/2}\right|^2}\sum\limits_{i,j,k} \frac{(B_y)_{i+1/2,j+1,k+1/2}^{n+1/2}-(B_y)_{i+1/2,j,k+1/2}^{n+1/2}}{\Delta y} S_{yy}^B+\nonumber \\
	+&\frac{b_{pz}^{n+1/2}}{\left|{\bf b}_p^{n+1/2}\right|^2}\sum\limits_{i,j,k} \frac{(B_z)_{i+1/2,j+1/2,k}^{n+1/2}-(B_z)_{i+1/2,j-1/2,k}^{n+1/2}}{\Delta y} S_{yz}^B,
\end{align}
where
\begin{align}
  S_{yx}^{B}=\sum\limits_{s\in n} &\left[\frac{Y_p^{s+1}-y_s}{Y_p^{n+1}-Y_p^n} S_1\left(y_{j}-\frac{y_s+Y_p^{s+1}}{2}\right)+\frac{y_s-Y_p^{s}}{Y_p^{n+1}-Y_p^n} S_1\left(y_{j}-\frac{y_s+Y_p^{s}}{2}\right)\right] S_{22}^{xz}[i,k+1/2], \\
	S_{yy}^{B}=\sum\limits_{s\in n} &\frac{Y_p^{s+1}-Y_p^s}{Y_p^{n+1}-Y_p^n} S_1\left(y_{j+1/2}-Y_p^{s+1/2}\right) S_{22}^{xz}[i+1/2,k+1/2], \\
	S_{yz}^{B}=\sum\limits_{s\in n} &\left[\frac{Y_p^{s+1}-y_s}{Y_p^{n+1}-Y_p^n} S_1\left(y_{j}-\frac{y_s+Y_p^{s+1}}{2}\right)+\frac{y_s-Y_p^{s}}{Y_p^{n+1}-Y_p^n} S_1\left(y_{j}-\frac{y_s+Y_p^{s}}{2}\right)\right] S_{22}^{xz}[i+1/2,k].
\end{align}
The $z$-component is obtained from the $x$-component by cyclic permutation of indices.

Let us now establish the rule for interpolating the magnetic field rotor from the grid to the particle, which is necessary for calculating the centrifugal drift. Note immediately that the energy balance imposes no restrictions on it: the vector ${\bf B}_p^{\ast n+1/2}$ is included in (\ref{drift_r}) and (\ref{vpar}) such that it completely cancels out when deriving (\ref{cond}), and the above chain of equalities is valid for any definition of $\left(\rot {\bf B}\right)_{p}^{n+1/2}$. Therefore, to find the interpolation rule, we require that $\left(\rot {\bf b}\right)_p^{n+1/2}$ vanishes in magnetic fields with straight lines. In particular, if the magnetic field has the form ${\bf B}=\left(0; 0; B_z(x,y)\right)$, then the rotor of the magnetic field must be associated with the gradient of its modulus in such a way as to ensure the exact equality $\left(\rot {\bf b}\right)_p^{n+1/2}=0$, that is
\begin{equation}
	\left(\rot {\bf B}\right)_{p}^{n+1/2}=\left[\left(\nabla B\right)_p^{n+1/2}\times {\bf b}_p^{n+1/2}\right].
\end{equation}
For example, the $x$-component of this equation gives
\begin{equation}
	\left(\rot {\bf B}\right)_{px}^{n+1/2}=\left(\frac{\partial B_z}{\partial y}\right)_p^{n+1/2}= b_{pz}^{n+1/2} \left(\nabla_y B\right)_p^{n+1/2}=\sum\limits_{i,j,k} \frac{(B_z)_{i+1/2,j+1/2,k}^{n+1/2}-(B_z)_{i+1/2,j-1/2,k}^{n+1/2}}{\Delta y} S_{yz}^B.
\end{equation}
By going through other directions of the guiding magnetic field and generalizing this rule to other components of $\left(\rot {\bf B}\right)_{p}^{n+1/2}$, we find that the interpolation of the magnetic field curl on a particle should be carried out using the non-diagonal components of the previously introduced tensor $S_{\alpha\beta}^B$:
\begin{align}
	&\left(\rot {\bf B}\right)_{p}^{n+1/2}=\nonumber \\
	&{\bf i} \sum\limits_{i,j,k} \left(\frac{(B_z)_{i+1/2,j+1/2,k}^{n+1/2}-(B_z)_{i+1/2,j-1/2,k}^{n+1/2}}{\Delta y} S_{yz}^B-\frac{(B_y)_{i+1/2,j,k+1/2}^{n+1/2}-(B_y)_{i+1/2,j,k-1/2}^{n+1/2}}{\Delta z} S_{zy}^B\right)+\nonumber \\
	&{\bf j} \sum\limits_{i,j,k} \left(\frac{(B_x)_{i,j+1/2,k+1/2}^{n+1/2}-(B_x)_{i,j+1/2,k-1/2}^{n+1/2}}{\Delta z} S_{zx}^B-\frac{(B_z)_{i+1/2,j+1/2,k}^{n+1/2}-(B_z)_{i-1/2,j+1/2,k}^{n+1/2}}{\Delta x} S_{xz}^B\right)+\nonumber \\
	&{\bf k} \sum\limits_{i,j,k} \left(\frac{(B_y)_{i+1/2,j,k+1/2}^{n+1/2}-(B_y)_{i-1/2,j,k+1/2}^{n+1/2}}{\Delta x} S_{xy}^B-\frac{(B_x)_{i,j+1/2,k+1/2}^{n+1/2}-(B_x)_{i,j-1/2,k+1/2}^{n+1/2}}{\Delta y} S_{yx}^B\right).\label{rotB}
\end{align}

\section{Numerical implementation}\label{s3}

The described drift-kinetic model is implemented as part of the general electromagnetic PIC code \texttt{xpic}~\cite{xpic}. Storage of grid electric and magnetic fields, as well as decomposition of the computational domain between processes, are provided by the PETSc (Portable Extensible Toolkit for Scientific Computation) library~\cite{petsc}.

The fully kinetic block of this code (Vlasov ions, the Crank-Nicolson energy-conserving scheme for their equations of motion, and the Maxwell equation solver) is described in detail in the paper~\cite{Kurshakov2026}.
Therefore, below we focus only on the implementation features of the drift-kinetic model: the structure of the time step with two nested iterative processes, the push of guiding centers, the consistent interpolation of $\nabla B$, $\rot\mathbf{B}$, and $\mathbf{M}$, and the trajectory segmentation. Elements common to both parts of the code are mentioned only to the extent necessary for understanding the drift-kinetic block.

The fully implicit approach we use couples the fields at the new layer to the trajectories of all particles, so the resulting system of finite-difference equations is nonlinear on two levels at once. The grid fields $\mathbf{E}_g^{n+1/2}$ and $\mathbf{B}_g^{n+1/2}$ enter the sources of Maxwell's equations through the current $\mathbf{J}_g^{n+1/2}$ and the magnetization $\mathbf{M}_g^{n+1/2}$, which are collected along the trajectories of the particles at the current step; the equations of motion for each guiding center contain the grid fields at the as-yet-undetermined point $\mathbf{R}_p^{n+1/2}$. The coupled system is solved by two nested iterative processes. The external process refines the fields $\mathbf{X}=(\mathbf{E}_g^{n+1/2},\mathbf{B}_g^{n+1/2})$ themselves using the nonlinear solver with Anderson-acceleration implemented in the PETSc library~\cite{Anderson1965}; the inner one solves the drift equations of motion for each guiding center using the Picard method. The Yee grid's rotor operators are formed once as sparse matrices (separately for the forward and backward differences corresponding to the relative shifts of the components $\mathbf{E}$ and $\mathbf{B}$) and then simply multiplied by the grid's field vectors. The key feature of this architecture is that the external solver does not access the particles directly: at each iteration, it requests only the current $\mathbf{J}_g^{n+1/2}$ and the magnetization $\mathbf{M}_g^{n+1/2}$, which are then collected by internally advancing the particles along the current field approximation. Self-consistency between the fields and particles is achieved precisely through this pair of quantities.

In the drift-kinetic approximation, a particle is identified with its gyrocenter, so its mechanical state is completely characterized by the gyrocenter's coordinate, its longitudinal velocity, and its magnetic moment $\mu_p$. Particles are distributed across grid cells; during the outer loop, information from the current and previous time layers is stored, allowing the particle pusher to be restarted from the state at layer $n$ at each outer iteration. Parallelization is based on a combination of MPI (Message Passing Interface) and OpenMP (Open Multi-Processing). The decomposition of the computational domain between MPI processes is inherited from the PETSc grid structures: each process owns its own grid block and the particles within it. The guiding center can leave a subdomain per step, so after updating coordinates, an exchange with neighbors occurs --- first with the numbers of outgoing particles, then with the particles themselves --- and their transitions across periodic boundaries are accounted for separately. Within a process, the particle advance cycle is distributed among OpenMP threads across cells, and the assembly of current and magnetization at nodes is performed using atomic additions, eliminating race conditions when multiple threads write simultaneously to shared nodes. Packing and unpacking of grid fields when passed to the solver is vectorized.

\subsection{Time integration}\label{imp:step}

By a time step, the state transits from layer $n$ to layer $n+1$. The current layer of particles is copied to the previous one (so that the values at layer $n$ remain available throughout all iterations), the initial approximation for the fields is taken from layer $n$, $\mathbf{X}^0=(\mathbf{E}_g^n,\mathbf{B}_g^n)$, after which the outer loop is launched. Upon its convergence, the fields are extrapolated to the integer layer, and particles that have changed cells or subdomains migrate between cells and processes. We first describe the inner loop, which the outer loop calls as a subroutine.

\paragraph{Inner loop: Picard method for a guiding center}
Let the grid field approximation be already chosen in the outer loop. Then the difference equations of motion~(\ref{drift_r}), (\ref{vpar}) are solved for the unknowns $\mathbf{x}_p=(\mathbf{R}_p^{n+1},\,v_{\|p}^{n+1})$ with a fixed magnetic moment $\mu_p$, which enters the system as a parameter. Since the right-hand sides depend on the field values at the point $\mathbf{R}_p^{n+1/2}=(\mathbf{R}_p^{n+1}+\mathbf{R}_p^n)/2$, which itself is determined by the sought $\mathbf{R}_p^{n+1}$, we write the system as a fixed point problem by analogy with the work~\cite{Chen2023}
\begin{equation}\label{pic_fp}
    \mathbf{x}_p^{k'+1}=\mathbf{G}\bigl(\mathbf{x}_p^{k'}\bigr),
\end{equation}
where $k'$ is the number of Picard iteration. Based on the current approximation $\mathbf{x}_p^{k'}$, half-step values $\mathbf{R}_p^{n+1/2,k'}$ and $v_{\|p}^{n+1/2,k'}$ are constructed; according to the interpolation rules (\ref{ef}), (\ref{bf}), (\ref{gradB}), (\ref{rotB}) with known endpoints of the segment $\mathbf{R}_p^n$ and $\mathbf{R}_p^{n+1,k'}$, the quantities $\mathbf{E}_p^{n+1/2,k'}$, $\mathbf{B}_p^{n+1/2,k'}$, $(\nabla B)_p^{n+1/2,k'}$ and $(\rot\mathbf{B})_p^{n+1/2,k'}$ acting on the particle are found; substitution into (\ref{drift_r}), (\ref{vpar}) gives the drift velocity $\mathbf{V}_p^{n+1/2,k'}$ and the longitudinal force $F_p^{n+1/2,k'}$. The next approximation is constructed using explicit formulas
\begin{align}
    \mathbf{R}_p^{n+1,k'+1} &= \mathbf{R}_p^n + \tau\,\mathbf{V}_p^{n+1/2,k'}, \label{pic_R}\\
    v_{\|p}^{n+1,k'+1}      &= v_{\|p}^n + \frac{\tau}{m_p}\,F_p^{n+1/2,k'}. \label{pic_v}
\end{align}
Convergence is controlled by the residuals of both equations of motion, related to the scale of the displacement per step,
\begin{align}
    \Bigl|\mathbf{R}_p^{n+1,k'+1}-\mathbf{R}_p^n-\tau\,\mathbf{V}_p^{n+1/2,k'+1}\Bigr|
        &< \varepsilon_a^{P} + \varepsilon_r^{P}\,\bigl|\tau\,\mathbf{V}_p^{n+1/2,k'+1}\bigr|, \label{pic_eR}\\
    \Bigl|v_{\|p}^{n+1,k'+1}-v_{\|p}^n-\tfrac{\tau}{m_p}\,F_p^{n+1/2,k'+1}\Bigr|
        &< \varepsilon_a^{P} + \varepsilon_r^{P}\,\bigl|\tfrac{\tau}{m_p}\,F_p^{n+1/2,k'+1}\bigr|. \label{pic_ev}
\end{align}
The relative criterion does not depend on the choice of the origin and the absolute position of the particle; the absolute threshold $\varepsilon_a^{P}$ only insures the degenerate case of zero shift. In the implementation, $\varepsilon_a^{P}=\varepsilon_r^{P}=10^{-12}$ and a maximum of $100$ iterations are assumed.

After convergence, the transverse velocity is restored from the conservation of magnetic moment,
\begin{equation}\label{pic_vp}
    v_{\perp p}^{n+1}=\sqrt{\frac{2\mu_p\,B_p^{n+1}}{m_p}},
\end{equation}
where $B_p^{n+1}$ is interpolated by rule~(\ref{bf}) to the point $\mathbf{R}_p^{n+1}$. The procedure is summarized in Algorithm~\ref{alg:picard}.

\begin{algorithm}[H]
\caption{Inner loop: Picard method for a guiding center}
\label{alg:picard}
\begin{algorithmic}[1]
\Require $\mathbf{R}_p^n$, $v_{\|p}^n$, $\mu_p$, $m_p$, $q_p$; approximation of fields $\mathbf{E}_g^{n+1/2}$, $\mathbf{B}_g^{n+1/2}$
\Ensure $\mathbf{R}_p^{n+1}$, $v_{\|p}^{n+1}$, $v_{\perp p}^{n+1}$
\State $\mathbf{R}_p^{n+1,0}\gets\mathbf{R}_p^n$, \enspace $v_{\|p}^{n+1,0}\gets v_{\|p}^n$
\For{$k'=0,1,\dots,\mathrm{maxit}$}
    \State build half-steps $\mathbf{R}_p^{n+1/2,k'}$, $v_{\|p}^{n+1/2,k'}$
    \State interpolate $\mathbf{E}_p^{n+1/2,k'}$, $\mathbf{B}_p^{n+1/2,k'}$, $(\nabla B)_p^{n+1/2,k'}$, $(\rot\mathbf{B})_p^{n+1/2,k'}$ \Comment{(\ref{ef}), (\ref{bf}), (\ref{gradB}), (\ref{rotB})}
    \State form $\mathbf{B}_p^{\ast n+1/2,k'}$ using~(\ref{bstar}); calculate $\mathbf{V}_p^{n+1/2,k'}$, $F_p^{n+1/2,k'}$ using~(\ref{drift_r}), (\ref{vpar})
    \If{criteria~(\ref{pic_eR}), (\ref{pic_ev}) are met} \State \textbf{exit the loop} \EndIf
    \State update $\mathbf{R}_p^{n+1,k'+1}$, $v_{\|p}^{n+1,k'+1}$ using~(\ref{pic_R}), (\ref{pic_v})
\EndFor
\State restore $v_{\perp p}^{n+1}$ using~(\ref{pic_vp})
\end{algorithmic}
\end{algorithm}

\paragraph{Outer loop: implicit system of Maxwell's equations}
The external process operates on grid fields at half-step --- vector $\mathbf{X}=(\mathbf{E}_g^{n+1/2},\mathbf{B}_g^{n+1/2})$ of dimension $6N_g$, where $N_g$ is the number of grid nodes. Using $f^{n+1}-f^n=2(f^{n+1/2}-f^n)$ and $\mathbf{H}=\mathbf{B}-\mathbf{M}$, finite-difference Maxwell's equations from Section \ref{s22} are rewritten as the equality to zero of the residuals
\begin{align}
    &\frac{\mathbf{E}_g^{n+1/2}-\mathbf{E}_g^n}{\tau/2}+\mathbf{J}_g^{n+1/2}(\mathbf{X})-\rot\mathbf{B}_g^{n+1/2}+\rot\mathbf{M}_g^{n+1/2}(\mathbf{X})=0, \label{mx_FE}\\
    &\frac{\mathbf{B}_g^{n+1/2}-\mathbf{B}_g^n}{\tau/2}+\rot\mathbf{E}_g^{n+1/2}=0, \label{mx_FB}
\end{align}
the left-hand sides of which are combined into a single vector-function $\mathbf{F}(\mathbf{X})$. The problem is reduced to finding its root $\mathbf{F}(\mathbf{X})=0$. The nonlinearity lies in the dependencies $\mathbf{J}_g^{n+1/2}(\mathbf{X})$ and $\mathbf{M}_g^{n+1/2}(\mathbf{X})$: for a given $\mathbf{X}$, the inner loop finds $\mathbf{R}_p^{n+1}$ and $v_{\|p}^{n+1}$ for each particle, after which the grid current and magnetization are collected from~(\ref{jf}) and~(\ref{magn}). Each calculation of $\mathbf{F}(\mathbf{X})$ costs one complete traversal of all particles --- this is the main cost of the external iteration, so the system is solved using the method of Anderson acceleration by simple iteration~\cite{Anderson1965}, implemented using the PETSc library~\cite{petsc} (without constructing the Jacobian). The iterations continue until the criterion is satisfied
\begin{equation}\label{mx_crit}
    \|\mathbf{F}(\mathbf{X}^k)\|_2 < \varepsilon_a^{A} + \varepsilon_r^{A}\,\|\mathbf{F}(\mathbf{X}^0)\|_2,
\end{equation}
where $\varepsilon_a^{A}$ and $\varepsilon_r^{A}$ are the absolute and relative tolerances of the external solver. In our implementation, we use $\varepsilon_a^{A}=\varepsilon_r^{A}=10^{-10}$, the maximum number of iterations 1000 and the initial approximation $\mathbf{X}^0=(\mathbf{E}_g^n,\mathbf{B}_g^n)$. Upon convergence, the fields are reconstructed on the integer layer using the half-sum rule:
\begin{equation}\label{mx_step}
    \mathbf{E}_g^{n+1}=2\mathbf{E}_g^{n+1/2}-\mathbf{E}_g^n,\qquad \mathbf{B}_g^{n+1}=2\mathbf{B}_g^{n+1/2}-\mathbf{B}_g^n .
\end{equation}
The inner Picard loop for a new approximation of $\mathbf{X}^k$ starts not with $(\mathbf{R}_p^n,v_{\|p}^n)$, but with $(\mathbf{R}_p^{n+1},v_{\|p}^{n+1})$, already found in the previous outer iteration: such a warm start refines the initial approximation and reduces the number of inner iterations. The outer loop is summarized in Algorithm~\ref{alg:anderson}.

\begin{algorithm}[H]
\caption{Outer loop (one time step)}
\label{alg:anderson}
\begin{algorithmic}[1]
\Require $\mathbf{E}_g^n$, $\mathbf{B}_g^n$; states of particles $\{(\mathbf{R}_p^n,v_{\|p}^n,\mu_p)\}$
\Ensure $\mathbf{E}_g^{n+1}$, $\mathbf{B}_g^{n+1}$; updated particle states
\State copy particle layer $n$ to the previous one; \enspace $\mathbf{X}^0\gets(\mathbf{E}_g^n,\mathbf{B}_g^n)$
\For{$k=0,1,\dots,\mathrm{maxit}$}
    \State for fields $\mathbf{X}^k$, for each particle, execute the Algorithm~\ref{alg:picard} $\rightarrow$ $\mathbf{R}_p^{n+1,k}$, $v_{\|p}^{n+1,k}$
    \State collect $\mathbf{J}_g^{n+1/2}(\mathbf{X}^k)$, $\mathbf{M}_g^{n+1/2}(\mathbf{X}^k)$ using~(\ref{jf}), (\ref{magn}) \Comment{Algorithm~\ref{alg:interp}}
    \State calculate the residual $\mathbf{F}(\mathbf{X}^k)$ using~(\ref{mx_FE}), (\ref{mx_FB})
    \If{criterion~(\ref{mx_crit}) is met} \State \textbf{exit the loop} \EndIf
    \State Anderson acceleration builds $\mathbf{X}^{k+1}$
\EndFor
\State extrapolate fields according to~(\ref{mx_step}); \enspace transfer particles across cells and subdomains
\end{algorithmic}
\end{algorithm}

The convergence of Anderson's method for the problem under consideration is not strictly proven, but is steadily observed in test simulations (section \ref{s4}).

\subsection{Field interpolation and trajectory segmentation}\label{imp:interp}

The exchange of fields and sources between particles and the mesh is implemented according to the rules derived in section \ref{s22}; the consistency of the interpolation functions is essential here.

The particle's straight trajectory at a given time step is cut into segments, each within its own cell. The partitioning planes are half-integer surfaces of the grid: a cell here is considered to be the volume constructed around the integer node $(i_s,j_s,k_s)$. It is under this assumption that rules~(\ref{gradB}) and~(\ref{rotB}) were derived. Particle transitions across the periodic boundary are processed before partitioning into cells, so that a segment crossing the domain boundary is stitched to its image on the opposite side.

The contributions of the segments to $\mathbf{E}_p$, $(\nabla B)_p$, $(\rot\mathbf{B})_p$ and to the grid current are summed with weights
\begin{equation}\label{seg_w}
	w_s=\frac{\ell_s}{\ell},\qquad \ell_s=\left|\mathbf{R}_p^{s+1}-\mathbf{R}_p^{s}\right|,\quad \ell=\sum\limits_{s}\ell_s=\left|\mathbf{R}_p^{n+1}-\mathbf{R}_p^{n}\right|,
\end{equation}
equal to the fraction of the segment length in the total trajectory length. For a rectilinear trajectory, the ratio of the lengths coincides with the ratio of the increments of any coordinate
\begin{equation}
	\frac{X_p^{s+1}-X_p^s}{X_p^{n+1}-X_p^n}=\frac{Y_p^{s+1}-Y_p^s}{Y_p^{n+1}-Y_p^n}=\frac{Z_p^{s+1}-Z_p^s}{Z_p^{n+1}-Z_p^n}=w_s,
\end{equation}
so ~(\ref{seg_w}) is exactly the weighting function that precedes the form factors in (\ref{ef}) and~(\ref{gradB}). In the degenerate case of an immobile  guiding center ($\ell\to0$), the trajectory is represented by a single segment with $w_s=1$, and the kernels continuously transit to ordinary interpolation $S_2$ at the point $\mathbf{R}_p^{n}$.

An additional fragmentation occurs in the kernels of $\nabla B$ and $\rot\mathbf{B}$: the form factor $S_1$, which appears when differentiating $S_2$ around a half-integer node, has a break at the integer node, and the segment containing this node within itself must be divided again with weights $a$ and $1-a$, where $a=(x_s-X_p^{s})/(X_p^{s+1}-X_p^{s})$ is the relative position of the node on the segment (see~(\ref{gradB})). If the node does not fall inside the segment ($a\notin(0;1)$) or the corresponding shift is zero, the division at $x_s$ is not performed and the form factor is taken at the segment midpoint. For large $\tau$, when a particle traverses several cells in a step, the scheme remains the same. The interpolation and assembly sequence for a single particle is given in Algorithm~\ref{alg:interp}.

\begin{algorithm}[H]
\caption{Field interpolation and source assembly for a single particle}
\label{alg:interp}
\begin{algorithmic}[1]
\Require ends of the trajectory $\mathbf{R}_p^n$, $\mathbf{R}_p^{n+1}$; $\mu_p$, $q_p$; grid fields $\mathbf{E}_g^{n+1/2}$, $\mathbf{B}_g^{n+1/2}$
\Ensure fields on the particle ($\mathbf{E}_p$, $\mathbf{B}_p$, $(\nabla B)_p$, $(\rot\mathbf{B})_p$); contributions to $\mathbf{J}_g^{n+1/2}$, $\mathbf{M}_g^{n+1/2}$
\State split $\mathbf{R}_p^n\to\mathbf{R}_p^{n+1}$ into segments within cells (taking into account periodic boundaries)
\State calculate $\mathbf{B}_p$ using form factor $S_2$ (\ref{bf}) in $\mathbf{R}_p^n$ and in $\mathbf{R}_p^{n+1}$
\State find $B_p^{n+1/2}$, ${\bf b}_p^{n+1/2}$ and ${\bf \hat{b}}_p^{n+1/2}$ using (\ref{bhalf})
\For{each segment $s$ with the weight $w_s=\ell_s/\ell$}
    \State add the contribution of $w_s$ to $\mathbf{E}_p$, $(\nabla B)_p$, $(\rot\mathbf{B})_p$ using (\ref{ef}), (\ref{gradB}), (\ref{rotB})
    \State decompose the current of the segment $\mathbf{J}_g^{n+1/2}$ using $\mathbf{\mathcal{S}}_{gp}$ (\ref{jf})
\EndFor
\State decompose magnetization $\mathbf{M}_g^{n+1/2}$ using (\ref{magn})
\end{algorithmic}
\end{algorithm}

A consistent choice of interpolation functions for $\mathbf{E}$, $\mathbf{J}$, $\nabla B$, $\rot\mathbf{B}$, and $\mathbf{M}$ turns the finite-difference equality~(\ref{cond}) into an identity for trajectory-averaged sources, so the energy balance~(\ref{kin_max}) holds with an accuracy no greater than the convergence accuracy of the external field solver. This statement is verified in Section \ref{s4}.

\section{Model testing}\label{s4}

The described scheme is benchmarked by a series of tests, each of which confirms the validity of the conclusions of Section \ref{s2}: the correct reproduction of the guiding center drifts, the finite-difference conservation of energy and charge, and the stability and convergence of the nested solvers for large steps of $\tau$ and $\Delta x$. It is convenient to begin with the dynamics of a single guiding center in given stationary fields --- such a test separates the errors of the drift-kinetic pusher from the errors of the field solver.

\subsection{Drift-kinetic pusher verification}

In all simulations in this subsection, we consider a single electron ($q_p=-1$, $m_p=1$) moving in the computational domain under externally applied constant fields ${\bf E}$ and ${\bf B}$. This electron does not create current, so it cannot influence on fields. By its initial coordinates and velocity, we calculate the position and velocity of its guiding center. The grid step is the same for all axes, $\Delta x=\Delta y=\Delta z=0.5\,c/\omega_{pe}$, the total computation time is $300\,\omega_{pe}^{-1}$. The drift-kinetic push of the particle is performed with a step $\tau=10\,\Omega_e^{-1}$, one and a half times greater than the Larmor rotation period. For clarity, we compare the guiding center trajectory with the trajectory of a real particle calculated by the fully kinetic pusher of \texttt{xpic}~\cite{Kurshakov2026} with a step $\tau=0.1\,\Omega_e^{-1}$ resolving the gyromotion. The drift velocity of the guiding center is compared with the analytical value predicted by equation~(\ref{drift_r}) for stationary fields.

\begin{figure}[p]%
	\centering
	\includegraphics[width=0.67\textwidth]{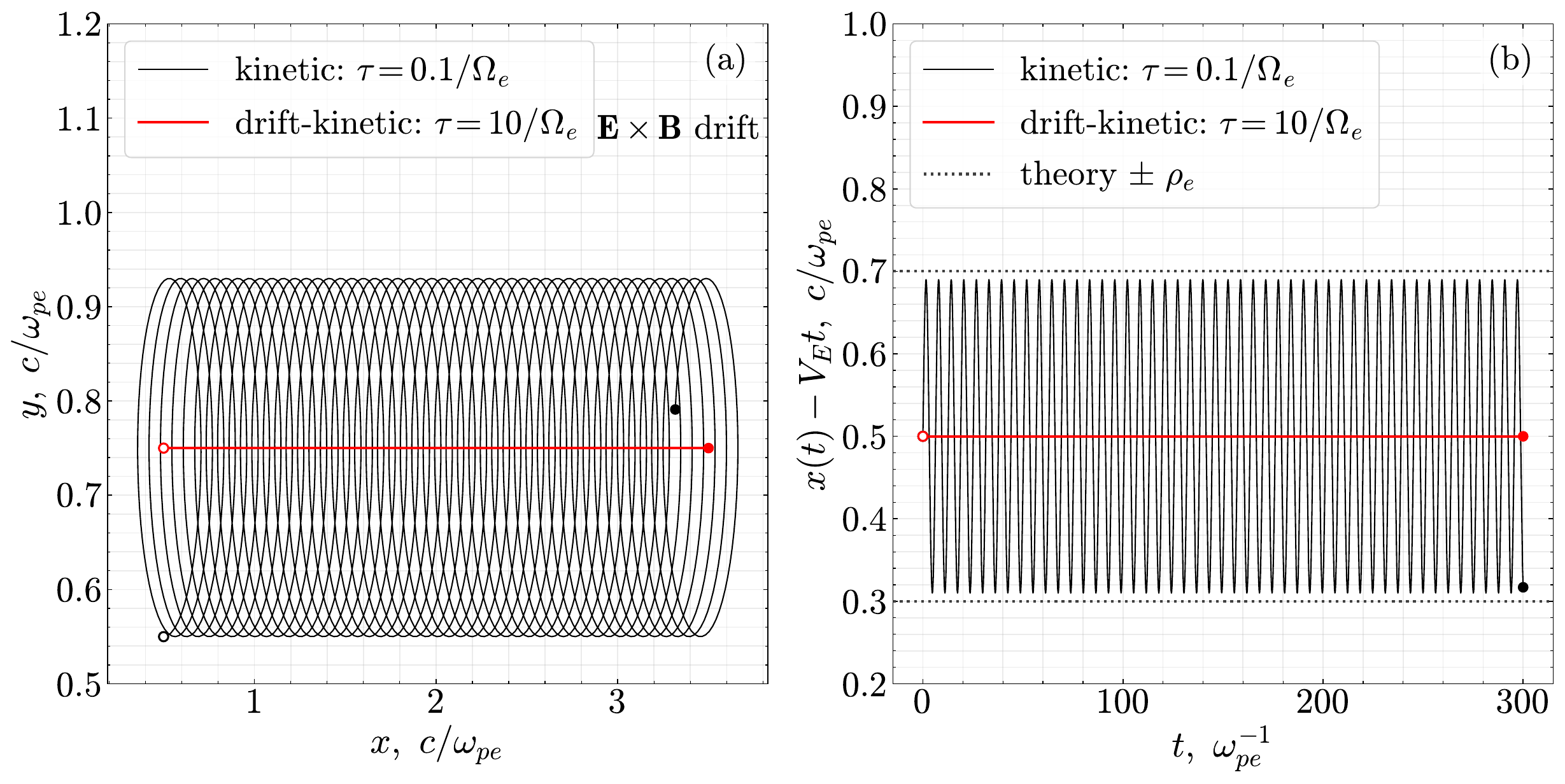}
    \caption{Electric drift of an electron in uniform fields ($\Delta x=0.5\,c/\omega_{pe}$, $\tau=10\,\Omega_e^{-1}$). (a) Trajectory in the $(x,y)$ plane: for the guiding center (red line) and for the real particle (black, $\tau=0.1\,\Omega_e^{-1}$). (b) Deviation from theory $x(t)-V_E t$; the dotted lines indicate the boundaries of $\pm\rho_e$. The hollow and filled circles mark the beginning and the end of the trajectory.}\label{fig:exb}

	\vspace{0.3em}
	\includegraphics[width=0.67\textwidth]{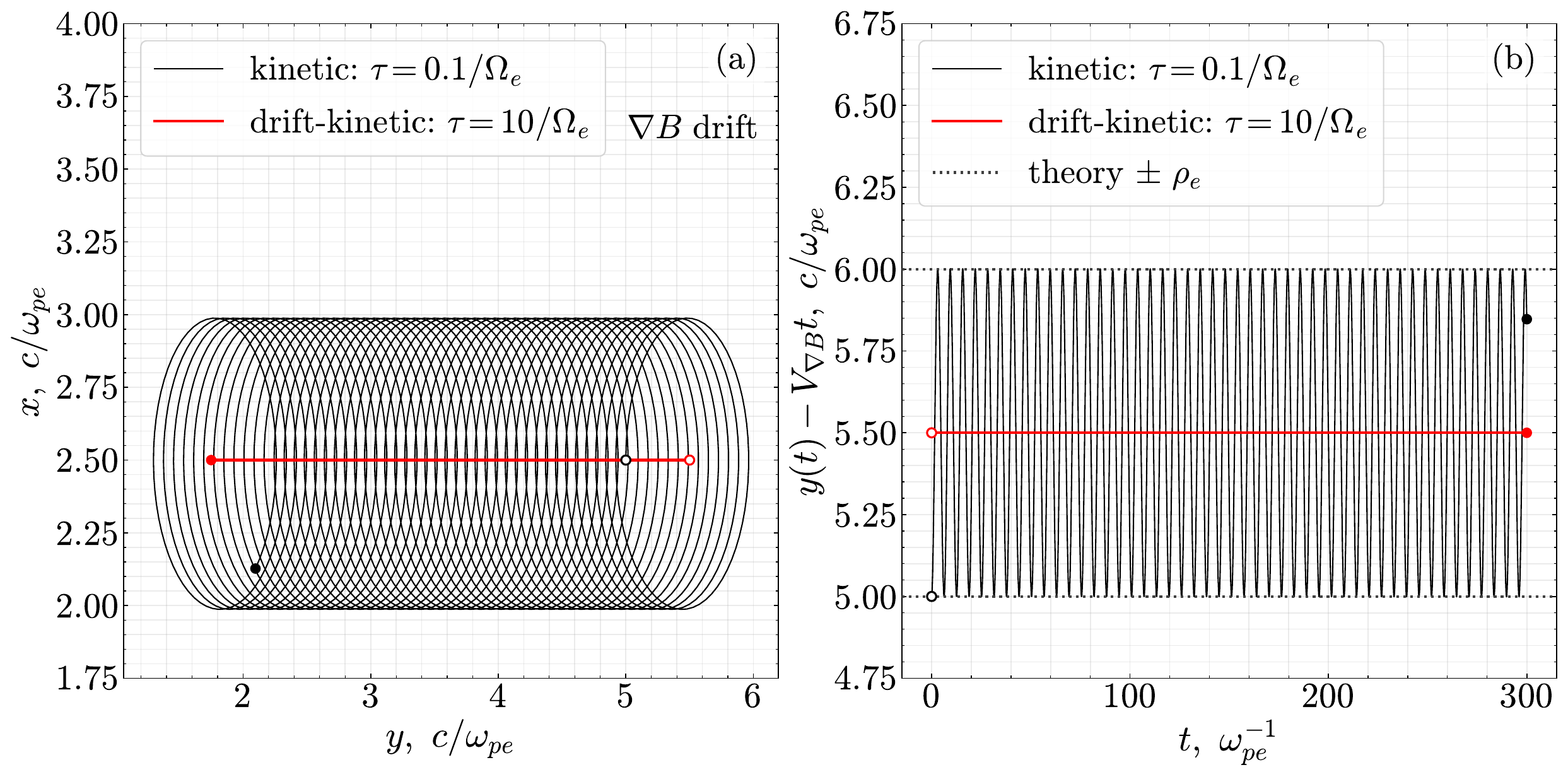}
	\caption{Magnetic gradient drift of an electron in the field of straight lines with a linearly increasing modulus ($\Delta x=0.5\,c/\omega_{pe}$, $\tau=10\,\Omega_e^{-1}$). (a) Trajectory in the $(y,x)$ plane for the guiding center (red line) and for the real particle (black, $\tau=0.1\,\Omega_e^{-1}$). (b) Deviation from theory $y(t)-V_{\nabla B}t$; the dotted lines indicate the boundaries of $\pm\rho_e$. Hollow and filled circles mark the beginning and the end of the trajectory.}\label{fig:gradb}

	\vspace{0.3em}
	\includegraphics[width=0.95\textwidth]{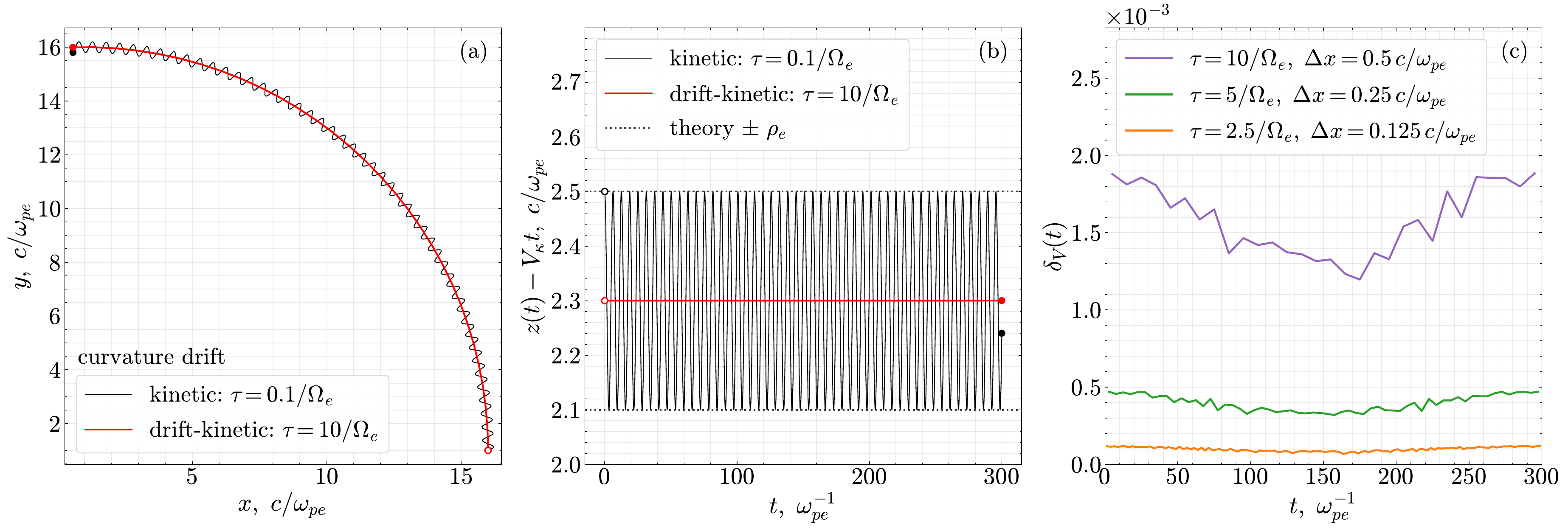}
		\caption{Centrifugal (curvature) drift of an electron in an azimuthal magnetic field of constant magnitude ($\Delta x=0.5\,c/\omega_{pe}$, $\tau=10\,\Omega_e^{-1}$). (a) Trajectory in the $(x,y)$ plane for the guiding center (red line) and for the real particle (black, $\tau=0.1\,\Omega_e^{-1}$); the hollow and filled circles mark the beginning and the end of the trajectory. (b) Deviation from theory $z(t)-V_\kappa t$; the dotted lines indicate the $\pm\rho_e$ boundaries. (c) Relative drift velocity error~(\ref{errV}) for three simulations with $(\tau,\Delta x)$ refinement from $(10\,\Omega_e^{-1},\,0.5\,c/\omega_{pe})$ to $(2.5\,\Omega_e^{-1},\,0.125\,c/\omega_{pe})$.}\label{fig:curv}
\end{figure}

The first test verifies the correctness of the electric drift. We set uniform fields ${\bf E}=0.01\,{\bf e}_y$ and ${\bf B}={\bf e}_z$, and launch the electron with the finite transverse $v_\perp=0.2$ and zero longitudinal velocity from the point $(0.5;\,0.55;\,0.75)\,c/\omega_{pe}$. The guiding center should move in the crossed fields with the velocity
 \begin{equation}\label{vexb}
	{\bf V}_E=\frac{\left[{\bf E}\times{\bf B}\right]}{B^2}=0.01\,{\bf e}_x
\end{equation}
independent of both the charge and the energy of the particle. Fig.~\ref{fig:exb}(a) shows that the red trajectory of the guiding center really lies on the axis of the well-resolved gyrotrajectory. Fig.~\ref{fig:exb}(b) shows how this motion looks in the reference frame moving with $V_E$: after subtracting the drift $x(t)-V_E t$, the real particle oscillates in the $\pm\rho_e$ band of the Larmor radius, while the guiding center stays at the same position. In uniform fields, there is no error associated with a finite size of a particle, so the numerical drift velocity coincides with $V_E$ down to machine precision.

The second simulation tests the magnetic gradient drift. To separate it from the centrifugal drift, we use the nonuniform magnetic field with straight field lines ${\bf B}=B_0\left(1+(x-x_0)/L\right){\bf e}_z$, where $B_0=1$ and $L=10$. Theoretical value for the corresponding drift velocity is
\begin{equation}\label{vgradb}
	{\bf V}_{\nabla B}=\frac{\mu_p}{q_p}\frac{\left[{\bf b}\times\nabla B\right]}{B}=\frac{\mu_p}{q_p}\frac{B_0/L}{B}\,{\bf e}_y.
\end{equation}
An electron with $v_\perp=0.5$ is placed at the point $(2.5;\,5.0;\,0.75)$, where $|{\bf B}|=B_0$. The linearity of $B(x)$ makes the finite-difference gradient $(B_{i+1}-B_i)/\Delta x$ identically equal to $B_0/L$, so the interpolation rule~(\ref{gradB}) reproduces the drift exactly; at a nonlinear profile the scheme would demonstrate the second-order accuracy in $\Delta x$. Figure~\ref{fig:gradb} is structured the same as the previous one: the guiding center moves along $y$ strictly with the velocity $V_{\nabla B}$, and subtracting this drift in panel (b) leaves only the gyration of the real particle in the $\pm\rho_e$ band.

The third test checks the interpolation of the magnetic field rotor. An azimuthal field of constant magnitude ${\bf B}=B_0\,{\bf e}_\varphi$ at $B_0=1$ with circular field lines around the $z$-axis yields $\nabla|{\bf B}|=0$, but $\rot{\bf b}={\bf e}_z/r$, so that longitudinal motion generates only the centrifugal drift along the $z$-axis:
\begin{equation}\label{vcurv}
	{\bf V}_\kappa=\frac{m_p v_\|^2}{q_p B_0 R}\,{\bf e}_z,
\end{equation}
where $R$ is the radius of the field line curvature, equal to the distance of the guiding center from the axis. An electron with $v_\perp=0.2$ and $v_\|=0.08$ is launched at a radius $R=15$. Unlike the two previous cases, the field here is nonlinear ($\sim1/r$), and the finite-difference curl is accurate only to the second order in $\Delta x$. We define the relative error of the drift velocity as
\begin{equation}\label{errV}
	\delta_V(t)=\left|\frac{V(t)-V_\kappa}{V_\kappa}\right|,\qquad V(t)=\frac{dZ_p}{dt},
\end{equation}
where $V(t)$ is the instantaneous velocity of the guiding center, and $V_\kappa$ is the theoretical value calculated from the initial parameters of the particle. Panels (a) and (b) in Fig.~\ref{fig:curv} repeat the scheme of the previous tests: the guiding center goes around a quarter circle (a) and drifts along $z$ with velocity $V_\kappa$ (b). Panel (c) collects $\delta_V(t)$ for three simulations with correlated refinement of the steps $(\tau,\Delta x)$. The error level decreases with grid refinement: $1.9\cdot10^{-3}$ for $\Delta x=0.5$, $4.7\cdot10^{-4}$ for $\Delta x=0.25$ and $1.2\cdot10^{-4}$ for $\Delta x=0.125$, which corresponds to the second order of convergence in $\Delta x$, embedded in the difference scheme.

In all three tests, the fields are static, so the total kinetic energy of the particle remains constant. The particle pusher ensures its conservation with machine-level precision over the entire calculation interval. The price of implicitness is moderate: in the most difficult problem (centrifugal drift), the internal Picard iteration converges in an average of 7 passes, compared to 2 for the fully kinetic pusher.

\subsection{Energy and charge conservation}

The following test shows that our proposed interpolation rules for fields, their gradients, and particle-generated currents allow us to satisfy conservation laws of global energy and local charge with the accuracy inherent in iterative solvers.  Let us study self-consistent evolution of an infinite uniform Maxwellian plasma immersed in a uniform magnetic field ${\bf B}={\bf e}_z$. In simulations of such a plasma, we use the cubic volume $30\times30\times30\,(c/\omega_{pe})^3$ with periodic boundary conditions. This volume is divided into $3\times3\times3$ cells with an intentionally large grid size $\Delta x=10\,c/\omega_{pe}\gg r_D$. The time step $\tau=10\,\Omega_e^{-1}\gg\omega_{pe}^{-1}$ exceeds the cyclotron period, and the total duration of the simulation reaches $t\approx8\times10^{4}\,\omega_{pe}^{-1}$ (8000 steps). The initial electric field is set equal to zero.
The plasma consists of electrons and ions, 15,000 of each type per cell, with equal densities of $n = 10^{13}$cm$^{-3}$ and temperatures of $T = 0.1$keV. For this test, we use a reduced mass ratio of $m_i/m_e = 100$ and describe both types of particles by the drift kinetics. The plasma has no flow velocity and remains in thermal equilibrium, so any increase in the total energy or charge residual should be considered as a discretization error.

The total energy of the periodic system does not contain a contribution associated with the Poynting flux, so its conservation law is reduced to the form
\begin{equation}\label{Wtot}
	W(t)=\sum\limits_g\left(\frac{|{\bf E}_g|^2}{2}+\frac{|{\bf B}_g|^2}{2}\right)+\sum\limits_p\left(m_p\frac{v_{\|p}^2}{2}+\mu_p B_p\right)=\mbox{\rm const}.
\end{equation}
We will control the accuracy of this law by calculating the relative energy error
\begin{equation}\label{deltW}
    \delta_W(t) = \frac{W(t)-W(0)}{W(0)}.
\end{equation}
Conservation of local charge in finite-difference form is controlled through the residual of the discrete continuity equation
\begin{equation}\label{charge_resid}
	\mathcal{R}_\rho(t) = \left(\sum_g\left(\frac{\rho_g^{n+1}-\rho_g^n}{\tau} + \left(\diverg{\bf J}\right)_g^{n+1/2}\right)^2\right)^{1/2},
\end{equation}
which vanishes identically if the Esirkepov's decomposition method applied to individual segments  is implemented correctly.

\begin{figure}[ht]
	\centering
	\includegraphics[width=\textwidth]{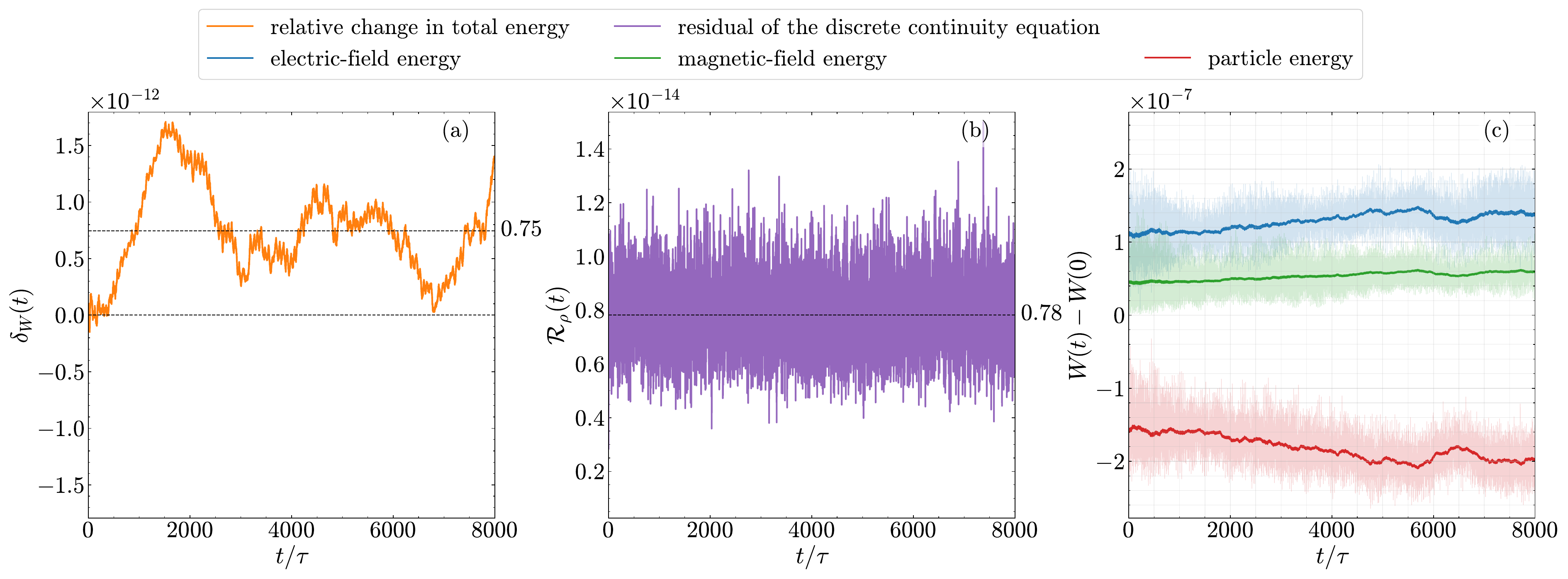}
    \caption{Conservation of energy and charge in a homogeneous Maxwellian plasma with a drift-kinetic description of particles ($\Delta x=10\,c/\omega_{pe}$, $\tau=10\,\Omega_e^{-1}$). (a) relative change in total energy $\delta_W(t)$~(\ref{deltW}); dashed lines --- zero level and the time-averaged value $7.5\times10^{-13}$. (b) discrepancy of a discrete continuity equation $\mathcal{R}_\rho(t)$~(\ref{charge_resid}); dashed line --- the time-averaged value $7.8\times10^{-15}$. (c) deviation of the kinetic energy of all particles (red), of the electric field energy (blue) and of the magnetic field energy (green) from their respective initial values; bold curves --- running average, shaded bands --- the raw data.}
\label{fig:energy}
\end{figure}

Fig.~\ref{fig:energy} confirms both conservative properties of our scheme.
The relative change in energy (a) does not exceed $1.7\times10^{-12}$ and does not exhibit secular growth: the curve changes the sign of the derivative several times  and, on average, remains at the level of $7.5\times10^{-13}$ (dashed line) that is determined by roundoff errors of our iterative solver. The residual of continuity equation (b) fluctuates around $7.8\times10^{-15}$ over the entire interval, remaining machine-precision noise. Panel (c) shows what contributions to the total energy change are made by the kinetic energy of all particles (red curve), the energy of the electric field (blue), and the magnetic field energy (green). It can be seen that after a very rapid process of thermal noise generation (within a few steps), the energy of electromagnetic fluctuations does not almost change, remaining many orders of magnitude smaller than the total kinetic energy of the particles. This indicates that our model does not suffer from any numerical instabilities that, under conditions of energy conservation, could facilitate mutual exchanges (up to equilibration) between particles and electromagnetic noise. Thus, we can conclude that our numerical scheme is stable at large time and grid steps ($\Delta x\gg r_D$ and $\tau\gg\omega_{pe}^{-1}$) and preserves the total energy with a given accuracy.

\subsection{Magnetic field expulsion by plasma}\label{tests:diamag}

In this test, we model the problem of establishing equilibrium for a plasma sheet with finite $\beta$ in an initially uniform magnetic field (the drift-kinetic approximation is used for all particles). The goal is to prove the adequacy of the drift-kinetic model for describing the diamagnetic effect, and therefore for studying plasma equilibria that can evolve to states in which the plasma pressure is a significant fraction of the magnetic field pressure.

\begin{figure}[H]
    \centering

    \includegraphics[width=0.55\textwidth]
        {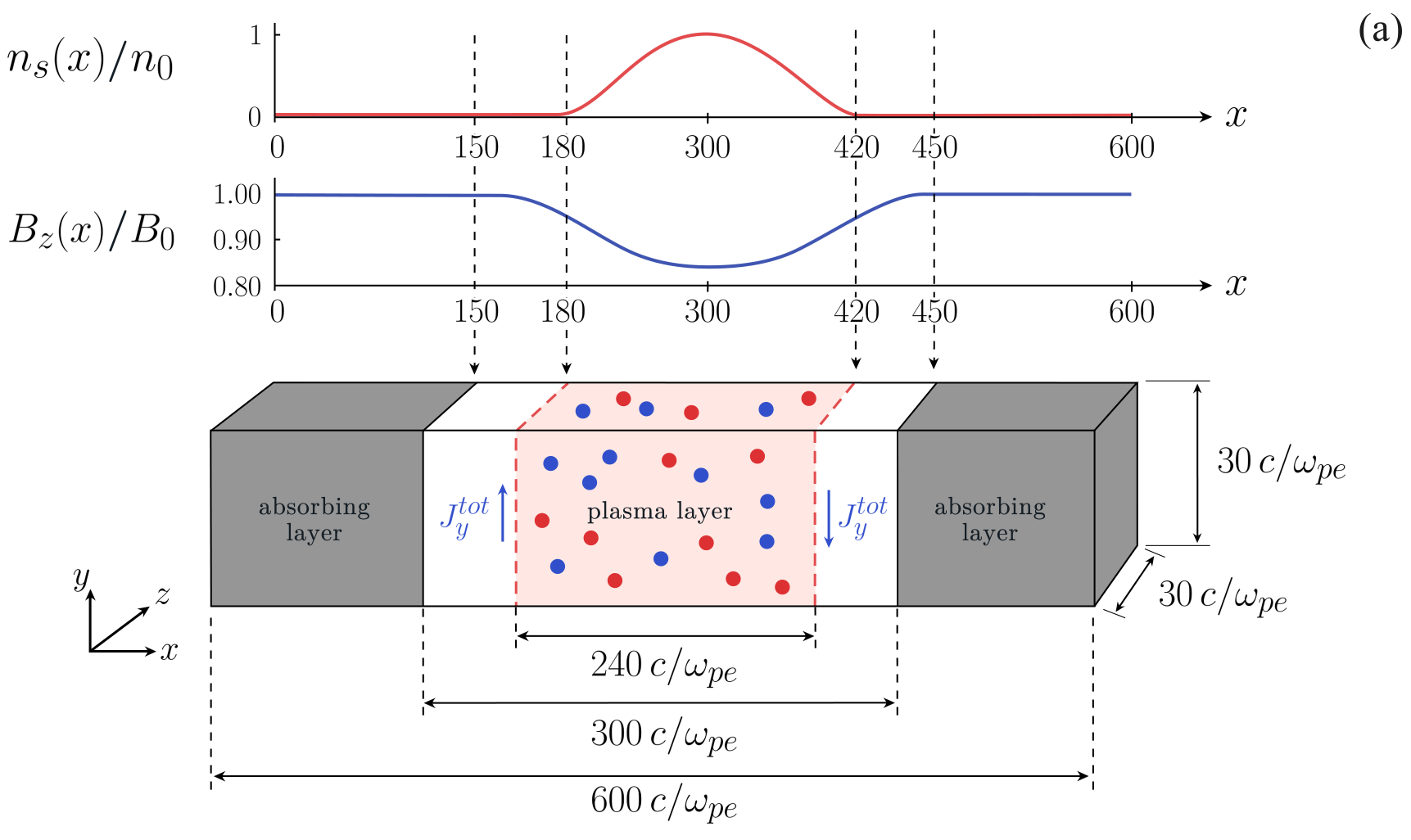}

    \vspace{3mm}

    \includegraphics[width=0.65\textwidth]
        {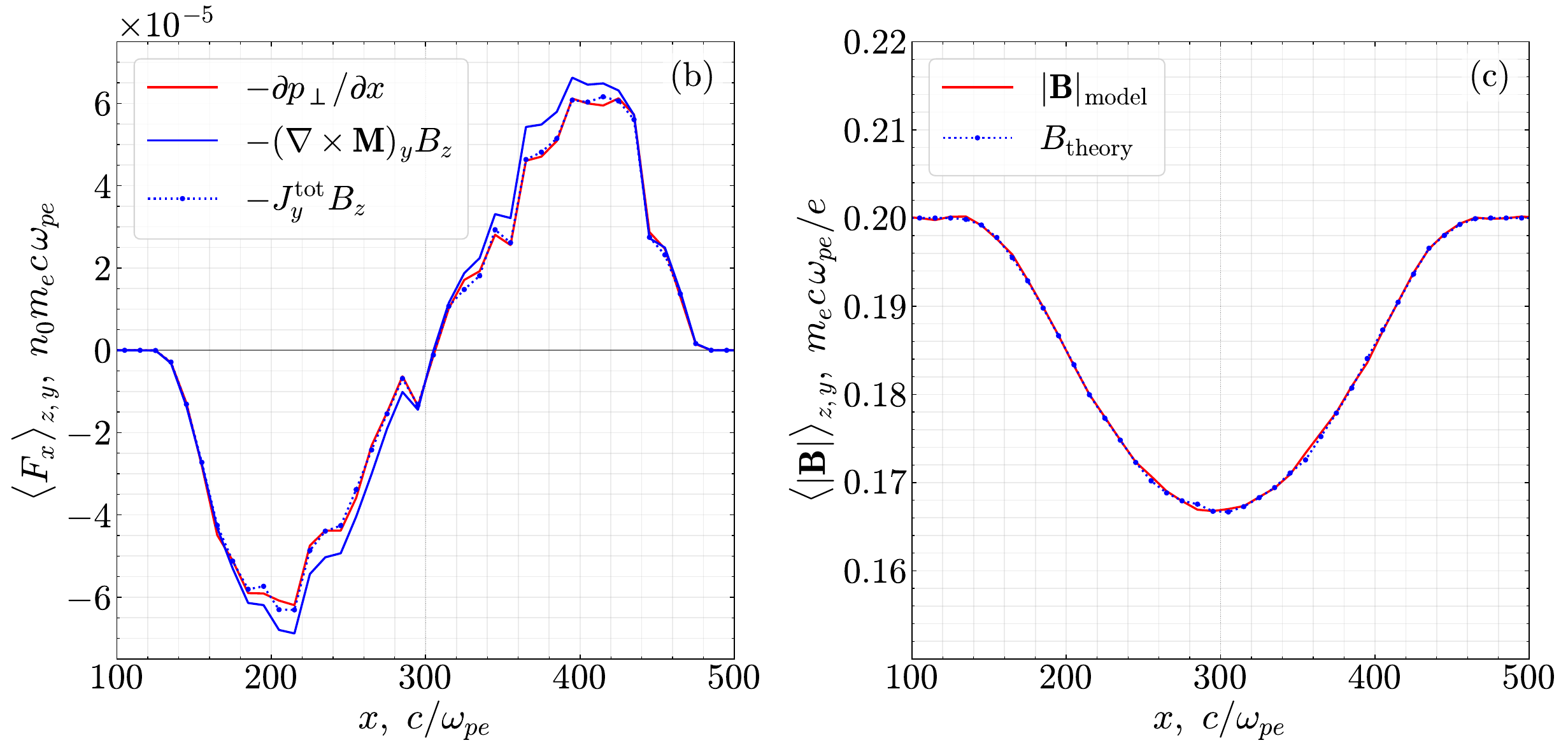}

    \caption{Test of magnetic field displacement by plasma ($\Delta x=10\,c/\omega_{pe}$, $\tau=10\,\Omega_e^{-1}$): (a) the layout of the computational domain with periodic boundary conditions along $y$ and $z$ (the absorbing layers are shown in grey and the initial plasma layer in pink), together with the initial density profile~(\ref{cosine_hump_1d}), normalized to its maximum value $n_0$, and the steady-state longitudinal magnetic-field profile $B_z(x)$, normalized to the external field $B_0$; (b) the force balance~(\ref{force_balance}) at $t=t_{\max}$ (pressure gradient $-dp_\perp/dx$ --- red curve, Ampere force due to the magnetization current $-(\rot{\bf M})_yB_z$ --- solid blue curve, and Ampere force due to the total current $-J_y^{\mathrm{tot}}B_z$ --- dotted blue curve); (c) comparison of the simulated magnetic-field profile $\langle B\rangle_{y,z}(x)$ (red curve) with the theoretical equilibrium profile~(\ref{p_balance}) (dotted blue curve).}
    \label{fig:diamag}
\end{figure}

The computational domain is a rectangular parallelepiped $L_x\times L_y\times L_z=600\times30\times30\,(c/\omega_{pe})^3$, divided by a grid into $60\times3\times3$ cells with steps $\Delta x=\Delta y=\Delta z=10\,c/\omega_{pe}$. Periodic boundaries are specified along the $y$ and $z$ axes, and along the $x$ axis there are two absorbing layers of width $150\,c/\omega_{pe}$ on each side, in which the fields are attenuated with a coefficient of $0.2$ per time step. The aim of these extended absorbing boundaries is to completely absorb the electromagnetic wave that is generated by the initial imbalance in pressure.  This geometry makes the problem quasi-1D -- all average values over $y$ and $z$ depend only on $x$. The initial magnetic field is uniform and directed along the $z$ axis, ${\bf B}_0=0.2\,{\bf e}_z$, and the initial electric field is zero. The layout of the computational domain is shown in Fig.~\ref{fig:diamag},\,(a).

The plasma consists of electrons and positrons which have unit mass, $m_e=m_p=1$, opposite charges $q_{e,p}=\mp1$ and Maxwellian velocity distributions with the same initial temperature $T_e=T_p=0.75$ keV. The particles are uniformly distributed along $y$ and $z$ and form a finite-size layer along $x$ ($x\in[180;\,420]\,c/\omega_{pe}$). The density inside this layer has the initial profile
\begin{equation}\label{cosine_hump_1d}
    n_s(x)=\frac{n_0}{2}\left[1+\cos\!\left(\frac{2\pi(x-x_c)}{L_x}\right)\right].
\end{equation}
where $s$ is the particle type ($e$ or $p$), $x_c=(x_{\max}+x_{\min})/2=300$ is the layer center coordinate, $L_x=x_{\max}-x_{\min}=240$ is its width, $n_0=3$ is the maximum density on the layer axis, and the average density over the layer is $\langle n\rangle=n_0/2=1.5$. The time step is $\tau=10\,\Omega_e^{-1}$, and the total computation time is $t_{\max}=5\times10^4\,\omega_{pe}^{-1}=10^4\,\Omega_e^{-1}$.
On average, $10^3$ macroparticles of each type per cell is used in this simulation.

In a steady state, the equation of transverse plasma equilibrium in a longitudinal magnetic field has the form
\begin{equation}\label{force_balance}
    \frac{dp_\perp}{dx}=J_y^{\text{tot}}B_z,
\end{equation}
where $p_\perp(x)=n_e(x)\,T_{e,\perp}+n_p(x)\,T_{p,\perp}$ is the transverse plasma pressure, $J_y^{\text{tot}}$ is the total current (its direction is schematically shown in Fig. \ref{fig:diamag}). The total current produced by both types of particles contains not only contribution of guiding centers, but also the magnetization current, which dominates in this balance:
\begin{equation}\label{Jtot}
    J_y^{\text{tot}}=J_y+\left(\rot{\bf M}\right)_y.
\end{equation}
From the stationary Maxwell's equation $\rot{\bf B}={\bf J}^{\text{tot}}$, we obtain $J_y^{\text{tot}}=-dB_z/dx$. Then, integrating~(\ref{force_balance}) with respect to $x$ yields the pressure balance:
\begin{equation}\label{p_balance}
    \frac{B_z^2(x)}{2}+p_\perp(x)=\frac{B_0^2}{2}.
\end{equation}

To check the balances~(\ref{force_balance}) and~(\ref{p_balance}), the $x$-profiles of pressure, current and magnetic field measured in the simulation are averaged over the transverse coordinates $y$ and $z$. Fig.~\ref{fig:diamag},\,(b),(c) shows how these balances are fulfilled at the moment $t_{max}$, when the transient process from the abrupt start (the fields at $t=0$ are not matched with the plasma) has already ended and the system evolution has achieved a steady state. In agreement with theory, the equilibrium current turns out to be less than the diamagnetic current by the value of the drift current produced by gyrocenters moving under the action of $\nabla B$.
In Fig.~\ref{fig:diamag},\,(c) one can see how strongly the magnetic field is expelled from the plasma:  in the center of the layer, the field drops from $B_0=0.2$ to $B_{\min}\approx0.168$, and, at the periphery, where plasma pressure tends to zero, it restores to the unperturbed value. The model profile $B_z(x)$ reproduces the theoretical curve~(\ref{p_balance}) with a maximum relative deviation of no more than $3\,\%$. In the force balance~(b), the difference between the Ampere force from the total current and the pressure gradient is of the same order.

The same simulation allows us to check the finite-difference energy balance in a setting where the system is no longer closed and macroscopic magnetic field gradients are present. The absorbing layers remove energy from the computational region, so the surface flux in the balance~(\ref{kin_max}) does not vanish, and the value of $W(t)$~(\ref{Wtot}) decreases. Its sum with losses remains constant, and the relative energy error is determined as
\begin{equation}\label{deltW_damp}
	\delta_W(t) = \frac{W(t)-W(0)+\Delta W_{\rm damp}(t)}{W(0)},
\end{equation}
where $\Delta W_{\rm damp}(t)$ is the total energy of the electromagnetic field absorbed by the boundaries since the beginning of the simulation.

\begin{figure}[ht]
	\centering
	\includegraphics[width=\textwidth]{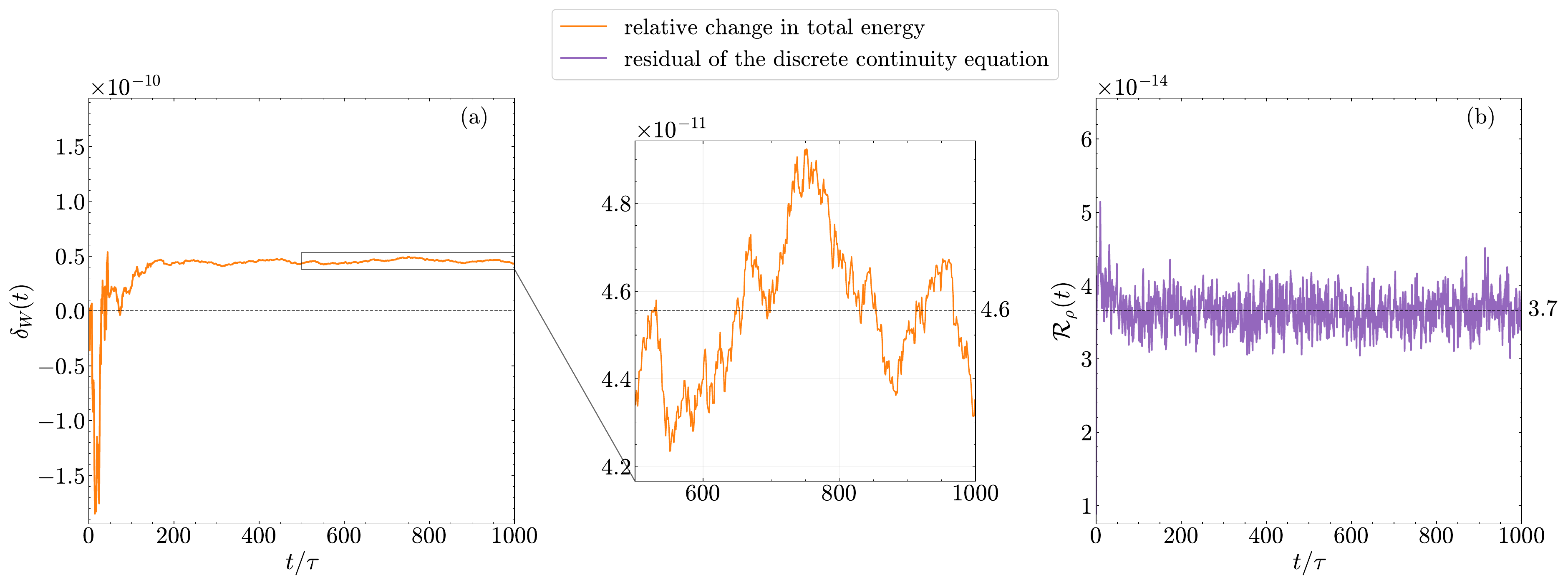}
	\caption{Conservation of energy and charge in the magnetic field displacement test ($\Delta x=10\,c/\omega_{pe}$, $\tau=10\,\Omega_e^{-1}$): (a) relative change in total energy taking into account the energy absorbed by the layers $\delta_W(t)$~(\ref{deltW_damp}); the enlarged fragment in the center shows the plateau stage $t/\tau\in[500;1000]$, where the dashed line marks the average over this interval, $4.6\times10^{-11}$; (b) discrepancy of a discrete continuity equation $\mathcal{R}_\rho(t)$~(\ref{charge_resid}); dashed line --- the average over the whole run, $3.7\times10^{-14}$.}
	\label{fig:diamag_energy}
\end{figure}

The result is shown in Fig.~\ref{fig:diamag_energy}.
The first $\sim200$ steps correspond to the transient process associated with the abrupt start. At this stage, the electromagnetic wave (generated by the initial imbalance between plasma pressure and magnetic field pressure) reaches the layers. By the time $t\approx200\,\tau$, the system reaches a stationary equilibrium, the energy ceases to lose in the absorbing layers, and the energy error curve (a) lies on a plateau of $4.6\times10^{-11}$. The plateau level is determined by the accuracy with which $\Delta W_{\rm damp}$ is measured: the error in this estimate accumulates only while there is a noticeable flux through the layers, that is, during the transient process, and then remains constant. The enlarged fragment in the center of the figure shows the second half of the simulation time, where $\delta_W$ oscillates in the band $(4.2\div4.9)\times10^{-11}$ without secular growth. The discrepancy of local charge conservation (b) throughout the simulation remains around $3.7\times10^{-14}$. Thus, neither the implementation of open boundary conditions nor the appearance of macroscopic magnetic field gradients led to a violation of the conservation laws embedded in the model.

\subsection{Ion-acoustic wave}\label{tests:sound}

The following test shows how accurately the drift-kinetic model for both electrons and ions reproduces the Landau damping of ion-acoustic wave traveling along the uniform magnetic field.
It is well known that the linear theory of longitudinal oscillations in a collisionless Maxwellian plasma yields the following dispersion relation for their frequency $\omega$ and wave number $k$:
\begin{equation}\label{ia_disp}
    \varepsilon_L(\omega,k)=1+\sum_{s=e,i}\frac{1}{k^2r_{Ds}^2}
    \Bigl[1+\zeta_s Z(\zeta_s)\Bigr]=0,
    \qquad \zeta_s=\frac{\omega}{\sqrt{2}\,k\,v_{Ts}},
\end{equation}
where $\zeta_s$ is the phase velocity of the wave normalized to the thermal velocity, and
\begin{equation}\label{ia_Z}
    Z(\zeta)=\frac{1}{\sqrt{\pi}}\int\limits_{\mathcal{C}}
    \frac{e^{-u^2}}{u-\zeta}\,du
\end{equation}
is the plasma dispersion function. The solution to this equation for a chosen value of $k$ is a complex quantity $\omega = \omega_s - i\Gamma_s$, where the real part determines the wave period $T = 2\pi/\omega_s$ and the imaginary part, $\Gamma_s$, determines the amplitude decay rate. Typically, to excite ion-acoustic oscillations in a non-isothermal ($T_e \gg T_i$) plasma of length $L$, an initial periodic density perturbation with wave number $k = 2\pi/L$ is created; this perturbation subsequently evolves as a standing wave. Monitoring the accuracy with which the wave damping rate is reproduced in such a problem is difficult, as the amplitude of the excited spatial harmonic oscillates at twice the frequency, meaning the amplitude decay can only be assessed by the reduction of the oscillation envelope. Furthermore, the simultaneous excitation of two counter-propagating harmonics creates favorable conditions for nonlinear interaction between them, which further complicates comparison with linear theory. To avoid these difficulties, we consider the damping of a traveling wave.

To compare the characteristics of a traveling ion-acoustic wave with theory in detail, we will solve the problem of the evolution of an ion-acoustic perturbation from a given initial state. We decompose the distribution function for particles of species $s$ into an unperturbed part and a small perturbation: $f_s(z,v_{\|},t)=F_{0s}(v_{\|})+\delta f_s(z,v_{\|},t)$, where $F_{0s}(v_{\|})=n_0(2\pi v_{Ts}^2)^{-1/2}\exp(-v_{\|}^2/2v_{Ts}^2)$ is the Maxwellian distribution in longitudinal velocity, normalized to the density $\int F_{0s}\,dv_{\|}=n_0$. The perturbations contain a single spatial harmonic with wavenumber $k$.
\begin{equation}
	\delta E_z=\hat E(t)\,e^{ikz}, \qquad \delta f_s=\hat f_s(v_{\|},t)\,e^{ikz}.
\end{equation}
Let the wave amplitude at the initial moment be characterized by the value $\hat E(0)=E_0$; then, from the linear response to a traveling mode $\propto e^{ikz-i\omega t}$, one can determine the required initial amplitudes of the perturbations in the distribution function, density, and flow velocity:
\begin{align}
	&\hat f_s(v_{\|},0)=-\,i\,\frac{q_s}{m_s}\,E_0\,\frac{\partial F_{0s}/\partial v_{\|}}{\omega-k v_{\|}}, \label{fd} \\
	&\hat n_s=\int\hat f_s\,dv_{\|}=-\,i\,\frac{q_sE_0}{m_s}\,\frac{n_0}{k\,v_{Ts}^2} \bigl[1+\zeta_sZ(\zeta_s)\bigr], \\
	&\hat u_s=\frac{1}{n_0}\int v_{\|}\hat f_s\,dv_{\|}=\frac{\omega\,\hat n_s}{k\,n_0}.
\end{align}
Thus, the initial density and particle velocity profiles must have the form:
\begin{equation}\label{ia_init}
    n_s(z,0)=n_0\bigl[1+a_s\sin(kz+\varphi_{n,s})\bigr],
    \qquad
    u_{\|s}(z,0)=C_{u,s}\sin(kz+\varphi_{u,s}),
\end{equation}
where $a_s=|\hat n_s|/n_0$ is the relative amplitude of the density perturbation, $C_{u,s}$ is the amplitude of the longitudinal velocity perturbation, and $\varphi_{n,s}$ and $\varphi_{u,s}$ are the phase shifts relative to the phase of the electric field wave.

The strictly exponential decay of the eigenmode is an approximate solution of the problem. To account for the ballistic contribution arising from the deformation of the initial distribution function due to the thermal expansion of particles, we must solve the Cauchy problem for the linearized system of Vlasov-Maxwell equations:
\begin{equation}\label{ia_ivp}
    \frac{\partial \hat f_s}{\partial t}=-i k v_{\|} \hat f_s
    -\frac{q_s}{m_s}\hat E(t)\frac{\partial F_{0s}}{\partial v_{\|}},
    \qquad
    ik\,\hat E(t)=\sum_s q_s\!\int \hat f_s\,dv_{\|}.
\end{equation}
Integrating these equations using the fourth-order Runge-Kutta method with the initial conditions described above makes it possible to compare the evolution of the specified ion-acoustic perturbation with the results of similar PIC simulations.

For the PIC model test, we select the following parameters. The computational domain has dimensions $L_x \times L_y \times L_z = 3 \times 3 \times 32 \, (c/\omega_{pe})^3$. Periodic boundary conditions are applied in all three directions. The time and spatial steps are set to $\tau = 5 \, \Omega_e^{-1}$ and $\Delta x = \Delta y = \Delta z = 1 \, c/\omega_{pe}$. Electrons are characterized by a charge $q_e = -1$, mass $m_e = 1$, and temperature $T_e = 5 \, \text{keV}$, while ions have a charge $q_i = +1$, mass $m_i = 25 \, m_e$, and temperature $T_i = 100 \, \text{eV}$; the unperturbed densities of both species are identical, $n_0 = 10^{13} \, \text{cm}^{-3}$. The external magnetic field is uniform and directed along the $z$-axis, ${\bf B} = {\bf e}_z$. With the chosen temperature, the grid spacing is an order of magnitude larger than the Debye radius.

 To initialize the sound wave, we distribute the particles in velocity and space according to the perturbations $\delta f_s$ (\ref{fd}) and set the initial electric field. The ion-acoustic wavelength matches the length of the domain; thus, the wave number is fixed at $k=2\pi/L_z=0.19635\,\omega_{pe}/c$. A numerical solution of the dispersion equation yields a frequency of $\omega_s=3.9304\times10^{-3}\,\omega_{pe}$ and a damping rate of $\Gamma_s=4.7291\times10^{-4}\,\omega_{pe}$. We set the amplitude of ion density oscillations in the acoustic wave to $a_i=0.03$, so the resulting electric field and velocity perturbations should be $E_0=5.638\times10^{-5}$ and $C_{u,i}=6.049\times10^{-4}\,c$. The electron density amplitude differs from the ion amplitude by $0.036\%$, and the phase shift between the species is $9.4\times10^{-5}$ rad --- it is precisely this deviation from quasineutrality that sustains the field $E_0$.

We will calculate the amplitude of the mode under investigation as the Fourier harmonic of the density perturbation $\eta$,
 \begin{equation}\label{dnc_def}
    \frac{\delta n_{i,1}(t)}{n_0}=\frac{2}{L_z}\int\limits_0^{L_z}
    \eta(t,z)\,e^{-ikz}\,dz,
    \qquad \eta=\frac{\langle n_i\rangle_{x,y}}{n_0}-1,
\end{equation}
and introduce the following quantity as a measure of parasitic (noise) density fluctuations:
\begin{equation}
	\delta n_{\rm res}=\sqrt{2\left\langle \eta^2\right\rangle_z-\left|\delta n_{i,1}\right|^2},
\end{equation}
where $\left\langle ...\right\rangle_z$ denotes averaging over the length of the region.

\begin{figure}[ht]
    \centering
    \includegraphics[width=\textwidth]{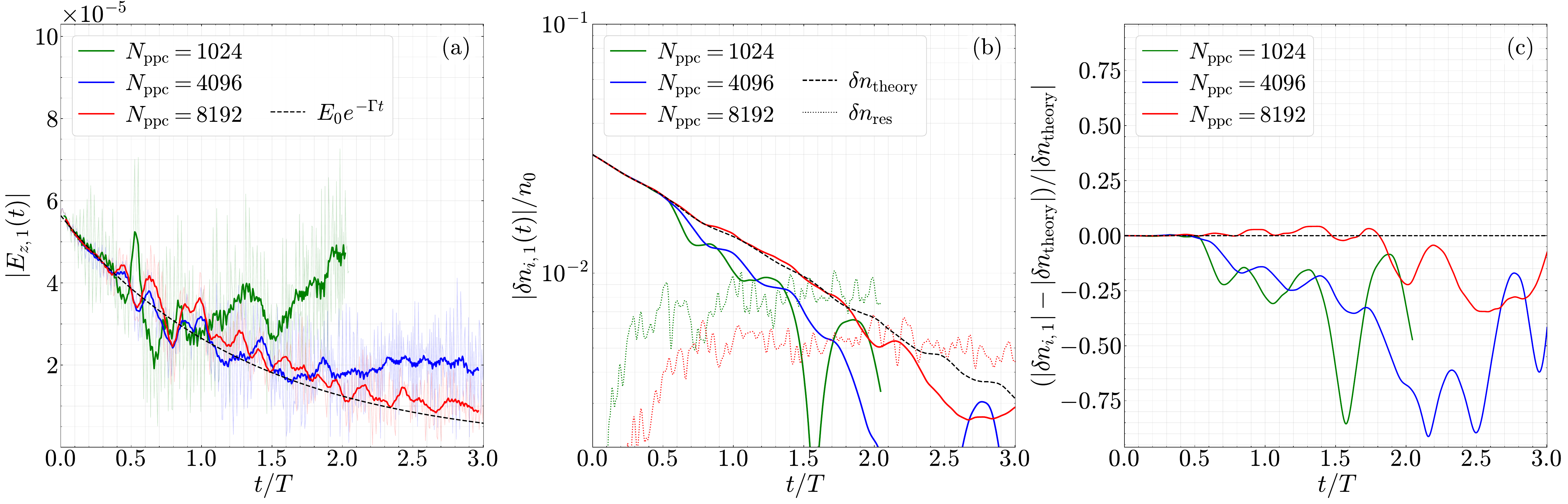}
    \caption{Landau damping of ion-acoustic wave ($\Delta z=1\,c/\omega_{pe}$, $\tau=5\,\Omega_e^{-1}$): (a) amplitude of the electric field harmonic $|E_{z,1}(t)|$ for different numbers $N_{\rm ppc}$ (bold curves --- running average over a sliding window of one wave period $T$, light curves --- the raw signal); black dashed line --- the theoretical decay $E_0e^{-\Gamma_s t}$; (b) amplitude of density perturbation $\delta n_{i,1}$ in an initially excited ion-acoustic mode (solid curves) and contribution to density perturbation from other harmonics $\delta n_{\rm res}$ (dotted, shown for $N_{\rm ppc}=1024$ and $8192$) for different numbers $N_{\rm ppc}$; black dashed line --- a numerical solution of the initial Cauchy problem~(\ref{ia_ivp}); (c) the relative deviation of $\delta n_{i,1}$ amplitude from the linear theory.}
    \label{fig:ia_wave}
\end{figure}

The results of PIC simulations performed for various numbers of particles per cell ($N_{\rm ppc}$) are presented in Fig. \ref{fig:ia_wave}.
Since the plasma remains nearly quasi-neutral during the acoustic oscillations, the electric field of the wave under study is extremely weak, making it difficult to quantitatively assess the agreement between its decay rate and theory due to the high noise. Fig. \ref{fig:ia_wave}(a) only shows that, with increasing particle number, the wave amplitude $E_{z,1}$ follows the theoretically predicted exponential decay of $e^{-\Gamma_s t}$ over increasingly longer time scales. A more detailed comparison with theory can be made by analyzing the evolution of much less noisy perturbations of the ion density. Fig. \ref{fig:ia_wave}(b) demonstrates that good agreement with theory (up to the stage where the wave amplitude drops to the thermal noise level $\delta n_{i,1} \approx \delta n_{\rm res}$) is achieved only with a sufficiently large number of particles per cell, $N_{\rm ppc} = 8192$ (red curve). Reducing this number leads to increased noise and more rapid energy loss from the selected mode. In each case, one can observe the stage where the wave amplitude begins to deviate sharply from the theoretical prediction; this is accompanied by the appearance of the second harmonic and is apparently caused by the onset of three-wave interactions. Up to these moments of time, the accuracy of agreement can be assessed from
Fig.~\ref{fig:ia_wave}(c) showing the relative deviation of the amplitude of a given mode
from the theoretically predicted value. It can be seen that in the simulation with
$N_{\rm ppc}=8192$ the relative error stays within $5\,\%$ up to $t\approx1.8\,T$, that
is over the whole stage preceding the sharp departure from the theoretical curve. As for the real part of the frequency $\omega_s$, its value is determined from the peak of the frequency Fourier spectrum and turns out to be 3\% higher than the theoretical prediction; given the spectral line width $\Gamma_s/\omega_s=0.12$, this can be considered good agreement.

\section{Conclusion}\label{s5}

In this work, an energy- and charge-conserving fully implicit PIC method is generalized to the case where the lightest particle species (electrons) is described within the framework of simplified drift kinetics. In this approximation, the motion of an individual particle is identified entirely with the motion of its guiding center, and the diamagnetic effect arising from the actual particle's cyclotron rotation is accounted for by assigning a magnetic moment to the gyrocenter. Given the large mass disparity between electrons and ions, we neglect the electric dipole moment of the electron gyrocenter and disregard the associated polarization current. Thus, to make an electromagnetic hybrid model with drift-kinetic electrons energy-conserving in continuous space-time, we should incorporate into the Maxwell equations  not only the current associated with gyrocenter motion but also the magnetization current $\nabla \times \mathbf{M}$.
The primary advantages of such a hybrid model are the ability to avoid resolving the electron cyclotron frequency and the capability to simulate the formation of equilibria characterized by the high relative plasma pressure values ($\beta < 0.6$) typical of open traps, without being restricted to considering small deviations from a prescribed equilibrium state.

To ensure that the fully implicit discretization of the differential equations governing the system's evolution does not violate global energy and local charge conservation laws, it was necessary to propose a method for the consistent calculation of $\nabla B$, $\nabla \times {\bf b}$, and the magnetization vector ${\bf M}$ --- quantities that determine the guiding-center velocity and diamagnetic current in drift kinetics. It turned out that if the interpolation of the magnetization vector from particles to the grid at the midpoint of the time step is performed using the symmetrized parabolic kernel $(S_2^{n+1}+S_2^{n})/2$, then the single constraint imposed by energy conservation can be reformulated as a rule for calculating all three components of the $\nabla B$ vector; this is made possible by a decomposition of the difference between form factors $S_2^{n+1}-S_2^{n}$ analogous to that used in the Esirkepov method.
As a result, the function $S_{\alpha\beta}^B$ for interpolating the magnetic field gradient from the standard Yee grid to the particle is structurally consistent with the kernel $\mathcal{S}_{gp}$ proposed in \cite{Chen2023}, but --- unlike that diagonal tensor --- it possesses non-zero off-diagonal elements. As for $\nabla \times \mathbf{b}$, its calculation requires components of the same tensor $S_{\alpha\beta}^B$.

To implement the proposed numerical scheme, a drift-kinetic module was integrated into the fully implicit electromagnetic PIC code~\cite{Kurshakov2026,xpic} that solves the full Vlasov equation. The nonlinear system of equations for guiding centers in evolving electromagnetic fields is solved using the PETSc library via two nested iterative processes: inner Picard iterations refine the gyrocenter trajectory with fixed grid fields, while an outer nonlinear Maxwell solver with Anderson acceleration refines the fields themselves, requesting only the current ${\bf J}_g^{n+1/2}$ and magnetization ${\bf M}_g^{n+1/2}$ from the inner loop.

The validity of the drift-kinetic model has been verified through a series of tests using large temporal and spatial steps ($\tau \gg \omega_{pe}^{-1}$ and $\Delta x \gg r_D$). The drift pusher accurately reproduces the analytical velocities of ${\bf E}\times{\bf B}$-, gradient, and centrifugal drifts for  a single particle in prescribed fields; furthermore, simulations with self-consistently evolving electromagnetic fields in a Maxwellian plasma confirm the conservation of energy and charge and show that errors determined by the solver's precision do not accumulated over time. It is demonstrated that, in a non-isothermal plasma, the numerical model correctly reproduces the Landau damping of an ion-acoustic wave traveling along the magnetic field. To verify the model's ability to describe the transition to finite-beta equilibria, a simulation was performed regarding the formation of a plasma layer with $\beta \approx 0.3$ within an initially uniform vacuum magnetic field. The results showed that, following the transient phase, the magnetic field is indeed displaced by plasma diamagnetic and $\nabla B$-drift currents, in good agreement with theory.

In the next stage, we plan to test a drift-kinetic model for electrons in a hybrid configuration with a fully kinetic ions. Subsequently, performance tests will be conducted, and the study of plasma confinement processes in open magnetic traps with strong mirrors will commence.

{\bf CRediT authorship contribution statement}

{\bf Oleg P. Morozov}: Investigation, Software, Validation, Visualization, Writing – original draft.

{\bf Igor V. Timofeev}: Conceptualization, Investigation, Formal analysis, Supervision, Validation, Writing – original draft.

{\bf Vladislav A. Kurshakov}: Investigation, Software, Validation.

\vspace{0.5cm}
{\bf Declaration of competing interest}

The authors declare that they have no known competing financial interests or personal relationships that could have appeared to influence the work reported in this paper.

\bibliographystyle{elsarticle-num}
\bibliography{ref}
\end{document}